\documentstyle[11pt,aaspp4]{article}
\newcommand{\kms}{\mbox{km s$^{-1}$}}
\newcommand{\etal}{{\it et al}.\ }

\newcommand{\bmv}{\mbox{B--V}}
\newcommand{\bmr}{\mbox{B--R}}
\newcommand{\mhi}{\mbox{${\cal M}_{HI}$}}
\newcommand{\HI}{\mbox{H\,{\sc i}}}

\newcommand{\HII}{\mbox{H\,{\sc ii}}}
\newcommand{\dark}{$\frac{{\cal M}_{HI}}{L_{V}}$}

\begin{document}

\title{B and V CCD Photometry of Southern, Extreme Late-Type Spiral Galaxies}

\author{Lynn D. Matthews}
\affil{Department of Physics \& Astronomy,
State University of New York at Stony Brook, Stony Brook, NY
11794-3800}
\affil{Electronic mail: matthews@gremlin.ess.sunysb.edu}
\authoraddr{Department of Earth and Space Sciences\\State University of 
New
York at Stony
Brook\\Stony Brook, NY 11794-2100}

\and
\author{John S. Gallagher, III}
\affil{Department of Astronomy, University of Wisconsin---Madison,
Madison, WI 53706}
\affil{Electronic mail: jsg@tiger.astro.wisc.edu}

%\slugcomment{Submitted to the Astronomical Journal}
                                                         
\begin{abstract}

We present B and V CCD aperture photometry for 
a morphologically-selected
sample of forty-nine southern, extreme late-type 
spiral galaxies.  All objects are moderate-to-low surface brightness
Local Supercluster field galaxies 
that were detected previously in \HI\ surveys.
Our sample features objects that have optical
luminosities, optical sizes, and \HI\ masses which are 
at the low end for spiral galaxies. These objects 
are not a new class of galaxy, but are examples of a common type of 
spiral galaxy 
that has been under-represented in nearby galaxy samples. 

We discuss the limitations of standard photometric
techniques as applied to extreme late-type spirals, and adopt a 
simple method of characterizing the light distributions 
of such objects in terms of aperture magnitudes, colors, color 
gradients, ``mean disk scales'', and optical structure.

In many 
respects, the global properties of extreme late-type spirals are intermediate 
between irregulars and ordinary high surface brightness
spiral galaxies. While our sample galaxies 
generally have 
regular structures and well-defined centers, some of the objects are 
very blue and/or have large gas fractions (\dark$\sim$3-9)---i.e. 
properties which have been traditionally associated with dwarf 
irregular galaxies.
Several galaxies become {\it redder} 
with increasing radius, a trend which seems to be uniquely found in
certain dwarf and very late-type disk galaxies. In addition, we find 
bars to be common in extreme late-type spirals. 

Our sample exhibits a diverse array of structural properties 
and morphologies in 
galaxies with otherwise similar physical parameters (e.g., M$_{B}$, 
\mhi, \bmv). A number
of the galaxies possess unresolved, point-like 
nuclei which may be related 
to low-luminosity AGNs or to M33-like compact nuclei. Finally, several 
of the galaxies show 
distinctive surface brightness ``steps'' in their disks, demonstrating 
that even small, late-type spirals can have multi-component disks.

\end{abstract}

\section{Background}
A number of recent studies have begun to explore sets of 
spiral galaxies which previously had remained largely absent from existing 
gaalxy catalogues due to observational biases. An important example is the low 
surface brightness
(LSB) spirals. Following 
the study of
Romanishin {\it et al.} (1982), several photometric studies of LSB 
spiral galaxies have appeared (e.g., Knezek 1993; McGaugh \& 
Bothun 1994; R\"onnback \& Bergvall 1994; Sprayberry {\it et al.} 1995b; 
de~Blok {\it et al.} 1995; Karachentseva {\it et al.} 1996; Vennik {\it 
et al.} 1996). These investigations have provided 
some of the first detailed photometric studies of LSB galaxies 
that are not dwarf irregular systems. They have brought attention to 
the fact that LSB disk 
galaxies span a wide range of Hubble types (e.g., Thuan \& Seitzer 
1979; Schombert {\it et
al.} 1992), although they are predominantly 
late-type spiral and irregular galaxies. 

Because LSB spirals tend to be gas-rich and appear ``underevolved'', 
studies of such galaxies have the potential  to 
aid in our understanding of  issues ranging from galaxy 
structure, formation, and evolution, to the origin of Lyman~$\alpha$
absorbers and the faint blue galaxy population (see, e.g.,  Rao \& Briggs 
1993; Rao \etal\ 1995; Impey 1993; van~der~Hulst {\it et al.}
1993; R\"onnback \& Bergvall 1994). However, in order to 
address these questions,  it is imperative to
explore in more detail the full range of properties spiral disks encompass.

With the exception of R\"onnback \& Bergvall (1994),
existing photometric
studies of LSB spirals 
have focussed on northern objects, selected by optical surface 
brightness  (e.g., McGaugh \& 
Bothun 1994; de~Blok {\it et al.} 1995) and/or within some 
range of \HI\ masses (e.g., Knezek 1993). Due to a combination of 
selection factors, these 
authors have emphasized large (e.g., Knezek 1993; Sprayberry {\it et 
al.} 1995b) or moderately large LSB spirals (e.g., de Blok {\it et al.} 
1995). In part this likely arises from
the presence of the Virgo Cluster and nearby galaxy voids in the 
northern hemisphere; because of these features,
northern field samples tend to be 
skewed toward either very nearby dwarf galaxies or 
larger, more luminous objects outside of the Local Supercluster. 
Few studies have focussed on populations of galaxies 
intermediate between recently 
discovered LSB spirals and existing samples of smaller, 
moderate-to-low 
surface brightness (MLSB) late-type galaxies within the Local 
Supercluster (e.g., de Vaucouleurs \etal\ 
1981; Longmore \etal\ 1982). 

In addition, it has been largely unrecognized that many small, MLSB
galaxies exhibit spiral structure (e.g., Matthews \& Gallagher 1993; 
van Zee \etal\ 1997;
but see also van~den~Bergh 1966; Longmore \etal\ 1982). 
This is partly because such objects 
were entirely missed due to their low surface brightnesses and/or small 
angular sizes, or because
the small, faint spirals which were part of existing samples were
mistakenly classified as irregulars due to the lack 
of high-quality optical images from which better morphological 
classifications could be made. As a result,  
MLSB Scd-Sm spirals largely have been excluded from
systematic studies of the spiral galaxy class even though such spirals 
may represent the most common type of gas-rich, organized, nearby disk 
galaxy (van der Kruit 1987).

In the current study we present photometry 
for such a sample of objects. Our sample includes spiral galaxies
which have mean disk surface brightnesses of
$\bar\mu_{B}\sim$23.3--25.6 mag arcsec$^{-2}$ and 
are which are, on average, physically smaller and/or less 
luminous than typical objects featured in other recent studies of 
LSB spirals. Hereafter we 
refer to these galaxies as {\it extreme late-type spirals}.

For the purpose of this study we define  ``extreme late-type 
spirals''\footnote{Commonly the term ``late-type spiral'' is used to 
encompass all spiral types Sbc or later, including small and giant 
spirals. 
To distinguish our sample objects as a special subset of 
late-type galaxies, we have adopted the nomenclature ``extreme 
late-type spirals''. We have avoided 
the use of the word ``dwarf spiral'', since there has been continual ambiguity 
in the literature as to whether this term implies some sort of 
luminosity cutoff, or instead denotes some set of structural properties 
(see Bingelli 1993). In addition, oftentimes
``dwarf'' is used to denote a galaxy that is not rotationally
supported (e.g, Tully \& Fouqu\'e 1985); however, ``dwarf spirals''
{\it are} rotationally supported disks.}
as ``the lowest-luminosity, late-type disk galaxies 
which still exhibit regular optical structures and centralized
light concentrations.'' In other words, these 
are Scd-Sm spirals which generally have mean surface brightnesses
below those of normal spirals (i.e. below the night sky level) and which
often fall at the low-end extreme for spiral galaxies in terms of 
properties such as \HI\ mass, optical luminosity, and optical size. 

Unlike other recent photometric studies of galaxies, we
focus exclusively on a {\it morphologically selected} 
sample of Local 
Supercluster, extreme late-type  spiral disks. By not 
applying a strict surface brightness, angular size, inclination, or 
\HI\ mass cutoff to our sample, 
we find we have selected a group of MLSB spirals which both complements 
and extends past studies. For example, unlike some existing 
photometric studies, we have not
excluded edge-on galaxies. Since dust obscuration in 
LSB galaxies may be low (e.g., 
R\"onnback \& Bergvall 1995; de~Blok {\it et al.} 1995; Kodaira \& 
Yamashita 1996), observed surface brightness will increase when LSB 
galaxies are viewed edge-on. Thus face-on samples might be biased
in favor of the selection of higher luminosity galaxies (which are 
easier to find and classify)
and against the selection of galaxies with the lowest 
intrinsic (i.e. inclination-corrected) central surface 
brightnesses, as such objects would become nearly invisible when 
viewed face-on. 

Our sample includes 
many galaxies which, prior to CCD imaging, would have been called ``irregular'' 
or ``peculiar'' rather than spirals (cf. Tables~1 \& 4), and includes some
spirals with lower luminosities and/or \HI\ masses  than the
spirals which typify recent LSB spiral samples (see e.g., Briggs 1997).  
Thus our sample helps to 
bridge the gap between past photometric studies of nearby Sm/Im systems 
such as the UK Schmidt photographic study of Longmore \etal\ (1982),
the DDO dwarfs (de~Vaucouleurs {\it et al.} 1981), extremely 
LSB Sm/Ims (R\"onnback \& Bergvall 1994) and larger LSB disks, which 
may be outside of the Local Supercluster.

As pointed out by McGaugh {\it et al.} (1995), morphological classifications
for MLSB galaxies are more difficult than for HSB objects
since features such as arm design (the main determinant of the 
assigned morphological type on the Hubble system) are not 
as pronounced in these
dim galaxies. However, the darker skies of the southern hemisphere 
offer an
extra advantage in discerning faint structures in MLSB objects; and based on 
their structural regularities and well-defined
centers, our sample galaxies cannot be classified as
irregulars. 
In addition, our sample galaxies tend to have pronounced radial 
surface brightness gradients, spiral-like \HI\ profiles (see Fouqu\'e 
{\it et al.} 1990; Gallagher {\it et al.} 1995; Matthews {\it et al.} 
1997b), and in several cases, point-like nuclei (see Section~5.1.4). All of 
these features help to confirm the spiral-like nature of these objects.
At the same
time, however, lack of evidence for a bulge component in most cases (see
Section~5.1.5) suggests that our galaxies are not early-type spirals (cf. 
Schombert {\it et al.} 1995).

In some ways our sample is heterogenous; it contains a range of 
arm designs, disk structures and thicknesses, and degrees of 
optical symmetry.  
Some of these spirals have low optical luminosities due to their 
very low surface brightnesses. Others have only moderately low
surface brightnesses, but are still ``underluminous'' due to their 
small sizes. What these disks have in common is that they represent 
the range of possible galaxy ``flavors'' among the smallest and least 
luminous organized disk galaxies. As we discuss below, one of the 
intriguing features which unites these galaxies is that many have
similar global physical properties in spite of their disparate optical 
morphologies.

\section{The New Extreme Late-Type Spiral Sample}
In total we present photometrically calibrated CCD images of 
forty-nine extreme late-type spiral field galaxies taken
through Johnson B
and V filters. The objects in our sample are listed in Table~1.
None of these objects 
had previously existing photoelectric or CCD B and V photometry or published 
\bmv\ colors. Our data 
were obtained as a part of a 
larger photometric survey (Gallagher \& Matthews, in preparation)
aimed at  southern, late-type, MLSB field galaxies  which had
been catalogued as extreme late-type spiral or irregular galaxies by 
Corwin \etal\ (1985) or Arp \& Madore (1987) and were subsequently
detected in \HI\ surveys of the Local Supercluster by
Fouqu\'e {\it et al.} (1990) or Gallagher {\it et al.} (1995).   

These \HI\ surveys 
uncovered many new moderately small \HI-mass,
Local Supercluster field galaxies
(\mhi$\sim$10$^{8}$--10$^{9}$M$_{\odot}$) at distances between 
10-40~Mpc. Previous \HI\ surveys had 
been largely incomplete in the
redshift range (V$_{r}<$3000~\kms) and declination band (-18$^{\circ} < 
\delta < -44^{\circ}$) of our study (Briggs 1997).
A few additional targets for our optical survey were chosen from 
Tully (1988) or Kraan-Korteweg \& Huchtmeier (1992);
these were galaxies with late-type classifications (T=8-10) and with
published \HI\ data but no optical B and V photometry. 

Our full imaging
survey revealed that a surprisingly diverse array of optical properties
exist among faint, moderately gas-rich, \HI-detected
nearby galaxies (e.g., Matthews \& 
Gallagher 1993). From these data we were able to cull the subsample
of objects having spiral-like properties that we present here.

\section{New Data Acquisition}
Our imaging data were obtained during the course of 7 nights at 
Cerro Tololo Inter American Observatory\footnote{Cerro Tololo 
Inter-American Observatory is operated by the Association of 
Universities for Research in Astronomy, Inc. under contract with the 
National Science Foundation.} in January 1993 using the 0.9m 
and 1.5m telescopes. All data were taken during dark time
under photometric conditions 
and reduced using the standard IRAF\footnote{IRAF is distributed by the
National Optical Astronomy Observatories, which is operated
by the Associated  Universities for Research in Astronomy, Inc.
under cooperative agreement with the National Science Foundation.} 
CCD reduction packages.  The 
mean sky brightness for each image is listed in Table~3.

\subsection{0.9m Observations}
We observed for 4 nights with the 0.9m f/13.5 telescope and Tek 512 
CCD with the VEB controller. CCD readnoise was $\sim$7.7~e$^{-}$ and the gain 
was 4.28~e$^{-}$/ADU. Our plate scale was 
0.445$''$ per pixel, and typical seeing was 0.7--1.3$''$. Due to our fairly 
small field of view (3.8$'\times$3.8$'$)
we focussed on galaxies whose catalogued diameters were 
less than 1.5$'$ for the 0.9m observations.   Exposure 
times were 600 seconds in B and 900 seconds in V.

The data were bias 
subtracted using a median of 25 bias frames taken each afternoon and were 
flatfielded by dividing by a median dome flat and by a median
twilight sky flat in the appropriate band. 

\subsection{1.5m Observations}
We observed for 3 nights with the 1.5m telescope, Tek~1024 CCD, and
VEB controller.  CCD readnoise was $\sim$3.8~e$^{-}$ and the gain was 
2.97~e$^{-}$/ADU.  We 
used the f/7.5 secondary which yielded a field of view of 7.2$'$ per 
side and a plate scale of 0.434$''$ per pixel.  

We found the counts in the overscan strip of the Tek~1024 to be, on 
average,
7 counts higher in our flatfield frames than in our object frames. Due 
to this apparent dependence of overscan level on total image counts, 
no overscan correction was applied.  A bias correction was performed 
by simply subtracting a median of 25 afternoon bias frames.

We found that small-scale structure was introduced by directly 
dividing by a median dome flat, so we produced a a modified dome 
flat by dividing the median afternoon dome flat by a 15$\times$15 pixel
median smoothed domeflat taken through the appropriate  filter.  Since our 
twilight sky  flats appeared stable from night-to-night, we averaged 
the twilight flats from the three nights together to increase 
signal-to-noise.  A 21$\times$21 pixel median smoothing was applied 
to remove stars, and finally the domeflat-corrected images were divided by 
this median-smoothed twilight sky frame.  
Exposure times for each galaxy were 500 seconds in both B and V.
Seeing ranged from 0.9-1.5$''$.

\subsection{Photometric Calibration}
Photometric solutions for our observing run were determined by monitoring 
several standard stars throughout each night. Aperture photometry was 
performed on the standard stars using the IRAF ``apphot'' package.
Since all nights were 
photometric, we combined all of our 0.9m standard star data to produce 
a global photometric transformation solution for all four nights. 
Similarly, we combined the three nights of 1.5m data to solve for 
a single solution. We used linear transformation equations from which 
we obtained zero points, color terms, and extinction coefficients for 
the data from each telescope. 
Only the \bmv\ transformation had an appreciable color 
term. Most of the scatter in the global solution comes 
from variable atmospheric extinction, as evidenced by 
the night-to-night variation 
in the derived extinction coefficient. Since 
we believe that extinction 
variations throughout each night and in different regions of the sky 
were comparable to the scatter in the night-to-night solutions, we have 
adopted the mean extinction coefficients from our global solution.
Our transformation equations are 
presented in Table~2. 

\subsection{Galaxy Photometry}
\subsubsection{The Problems of Surface Photometry of Late-Type and Low 
Surface Brightness Galaxies}
Accurate galaxy  surface photometry is a 
complicated problem both in practice [e.g., choices of model 
parameters, dependence of derived parameters on signal-to-noise of 
data (e.g., de Vaucouleurs 1977)] 
and in terms of one's ability to 
derive a characterization of a galaxy's optical light which can be 
interpreted in a physically meaningful manner (e.g., in cases where the 
galaxy has non-elliptical isophotes or a significant asymmetry). 

Traditionally, galaxy surface photometry involves fitting a series
of elliptical isophotes to an image and from this extracting
a magnitude and surface brightness profile based on the azimuthally
averaged fit (e.g., Kodaira \etal\ 1990). Several authors have 
pointed out the shortcomings of 
this technique for characterizing a galaxy's true nature (e.g., 
Burstein 1979; Kent 1984,1985; Odewahn 1991; Tully {\it et al.} 1996). 
Moreover, the quantities derived from functional
fits to the derived mean radial surface brightness profiles 
vary considerably between observers (Knapen \& van der Kruit 1991).

We found multiple isophotal fits to be largely unsuitable for 
efficiently photometering the types of 
galaxies in our present sample. Instead we adopt a 
simple photometric approach based on simulated multiple aperture photometry
and present a collection of optical measures of extreme late-type spiral
galaxies which are free from model-dependent assumptions.
We also provide comments on interesting features of 
individual objects in order that our sample can be used to select 
galaxies for follow-up studies, where more more detailed examinations 
of the light distributions can be made (e.g., to complement 2-D 
kinematic data).

\subsubsection{Photometry of the Extreme Late-Type Sample: The 
Procedure}
Our basic approach to photometering our galaxies was to use the 
IRAF/STSDAS ``ellipse'' package to fit a series of three concentric
elliptical apertures to each object. The outmost ellipse allows us to 
determine a limiting isophotal magnitude, inclination, position angle,
and axial ratio. 
The two additional ellipses provide a characterization of 
the light distribution of the galaxy in the nuclear region and a
middle disk region, plus a coarse measure of the color gradient 
and the steepness of the surface brightness profile
across the disk. 
Following, we outline the steps of 
our procedure and discuss our motivation for this 
approach.

\begin{enumerate}
\item A mean sky value for each B and V
frame was determined by measuring the mean 
value in 3-6  rectangular regions, 15 or more pixels on a side. 
These regions were chosen as near as possible 
to the center of the chip while still avoiding contamination from stars 
or the diffuse 
light of the galaxy. The average of these measurements was 
subtracted from each respective frame. 

\item Foreground stars and cosmic rays within the galaxy aperture
were removed by replacing circular regions of an 
appropriate radius with an interpolation of the surrounding pixels as 
measured in an annulus 
surrounding the masked region. We used H$\alpha$--R frames of the 
galaxies when available (Matthews \& 
Gallagher, unpublished) to distinguish foreground stars from 
\HII\ regions within the galaxy.
In a few images where the galaxy was contaminated by 
very bright stars with large diffraction spikes, complete removal of the 
starlight was difficult and photometric 
uncertainties are therefore larger; we have noted 
these cases in Table~3. 

\item The center of each galaxy was found by simply locating the pixels 
of maximum light concentration. This location generally could be 
unambiguously located to within 2-3 pixels (i.e., the
typical size of our seeing disk). 

\item The length of the semi-major axis and its position angle were 
measured on each V frame, and an ellipticity was estimated from 
the measured major to minor axis ratio. 

\item Using the estimated ellipse parameters, the outmost ellipse was 
fit to the galaxy.  (The V-band surface brightness of the outermost fitted 
ellipse for each galaxy is listed in Table~3). Small adjustments were 
made to the ellipse 
parameters until a best-fit ellipse was found to the outer galaxy 
isophote. The validity of these fits were verified on several 
independent passes through the data in which the data were displayed 
using alternate scalings and color tables 
to emphasize faint details. For some of the faintest objects we 
smoothed the images and again checked by eye the goodness of our outer 
ellipse fits.

\item Once the outermost ellipse parameters were established, two 
additional ellipses were fitted at $\frac{a}{1.8}$ and 
$\frac{a}{(1.8)^{2}}$, where ``$a$'' is the 
galaxy semi-major axis.   Position angle and ellipticity were held 
fixed. No effort was made to fit 
surface brightness ``steps'' (see Section~5.1.3) or features such 
as bars.  Instead we focus 
our attention on the global photometric properties of each galaxy. 
\end{enumerate}

The measured optical parameters for our sample galaxies are given in 
Table~3. Apparent magnitudes do not include correction for Galactic 
extinction (although extinction values are listed in Table~1).
We made no attempt to correct our derived magnitudes for internal 
extinction effects since these corrections are poorly know for MLSB galaxies. 
Based on analogies with other LSB galaxies, we believe that such corrections 
would be small for most of our sample due to 
low metallicities and dust contents (e.g., McGaugh 1994; 
R\"onnback \& Bergvall 1994; Kodaira \& Yamashita 1996).  We 
derived our inclinations based on the formula
\centerline{cos$^{2}$(i)=$\frac{(\frac{b}{a}) - q^{2}_{\circ}}{1 - 
q^{2}_{\circ}}$}

\noindent where $q_{\circ}$ for each object
was assigned based on the object's Hubble type according to Heidmann 
\etal\ (1972) and $\frac{b}{a}$ was determined from our outermost
elliptical aperture. The high frequency of optical asymmetries in 
extreme late-type spirals makes inclination derivations subject to more 
uncertainty than for HSB galaxies (see below) 
although the severity of this effect is difficult to quantify.

\subsubsection{Justification of Photometry Approach}
The photometry
method we describe allowed us to produce a set of useful photometric measures 
in an efficient manner without model-dependent 
assumptions. For example, azimuthal averaging of the radial
light distribution may facilitate the fitting of an exponential profile 
to the galaxy's disk, but if features like asymmetries or off-central light 
concentration are 
ignored, valuable physical 
information about the galaxy is thrown away (e.g., Burstein 1979; Ohta 
{\it et al.} 1990; Odewahn 1991) and derived results can be misleading 
(e.g., Richter \& Sancisi 1994). 
An examination of our galaxy images (Plates~1-7) quickly reveals
that
``irregularities'' seem  the rule rather than the exception for extreme 
late-type spirals. Thus it is especially important to consider this 
information in characterizing the optical properties of these galaxies. 

To illustrate the severity of these effects, we present a full 
isophotal ellipse fit for three of our sample galaxies
in Figure~1.  It is clear that while in Panel~a, the light profile of 
ESO~548-050 can be
adequately fit  by a single exponential, this characterization is 
deceptively simple since an examination of the galaxy image (Plate~4),
clearly reveals a significant outer disk asymmetry. Panels
b and c (ESO~358-015 and ESO~358-020 respectively)
show cases where a single exponential  does not
characterize the disk's light distribution, and multiple disk 
components are required (neither galaxy has a discernible bulge or bar
on the optical images). Here the definition of ``disk scale length''
as defined by a single exponential fit becomes ambiguous.

Radial surface brightness profiles of LSBs in other recent papers
also show deviations from single exponentials in galaxies
which do not possess bulge components (e.g., Odewahn 1991;
R\"onnback \& Bergvall 1994). The deviations become especially
striking when compared with high-quality surface brightness profiles
of HSB galaxies, which are often very well fit with single exponentials
(plus a bulge component; e.g., de Jong 1995, but cf. Boronson
1981).

For extreme late-type spirals, the motivation for a simple 
photometry approach is strengthened by practical concerns. Because 
images of MLSB 
galaxies are inherently of relatively low signal-to-noise, one must be 
cautious about attempting automated fitting procedures.
For our data we found that the IRAF/STSDAS ``ellipse'' program was 
incapable of making suitable ellipse fits to MLSB
objects in even a semi-automated manner and the program 
was unable to chose a unique solution, especially in the faint outer 
parts of the galaxy. These difficulties made 
automated ellipse 
fitting for individual galaxies unreliable if ellipse
parameters are allowed to freely vary.

\section{Sources of Photometric Uncertainty}
\subsection{The Present Dataset}
The 0.9m and 1.5m telescopes with the Tek 512 and 1024 CCDs 
are good choices for photometry of extreme late-type spirals 
since both telescopes 
have minimal scattered light and both CCDs are cosmetically excellent 
with no detectable fringing. In addition, since most of our targets were field 
galaxies well out of the Galactic Plane, contamination from field stars 
was minimized. The combination of these factors allowed 
us to obtain flat fields, on average, to 
better than 2\% in B and 0.4\% in V with the 0.9m and 1.3\% in 
B and 
0.6\% in V with the 1.5m telescope. In most frames, flatfield uncertainty 
was the dominant source of photometric error. See the Appendix~1 for further 
details on our error estimation procedure.

As a check on the accuracy of our results, we compare our B magnitudes 
with the
photographic B magnitudes published in the ESO Catalogue (Lauberts \& 
Valentijn 1989) for 45 of our galaxies. The results are plotted
in Figure~2. An offset of $\sim+$.2 magnitudes
is evident for our new results. Such an offset is consistent with 
that found by other workers (e.g., R\"onnback \& Bergvall 1994; Vader 
\& Chaboyer 1994; Vennik \etal\ 1996) and verifies the basic 
integrity of our results. We note however, that the mean offset for our 
Night 6 data (0.287$\pm$.186 magnitudes; 7 galaxies) is larger than the 
mean of the other five nights (0.047$\pm$.32
magnitudes; 36 galaxies), although the scatter is large. We have no 
evidence that conditions were non-photometric on 
Night~6; 
however, a zero-point offset may be be present on that night. 
We have flagged these data in Table~3.

\subsection{The Problems of Determining Total Magnitudes of LSB Galaxies}
The conversion of a measured magnitude to a ``total'' magnitude for a 
galaxy is always fraught with uncertainties (e.g., de Vaucouleurs \& 
Corwin 1977). Unfortunately 
there arises an additional set of concerns for MLSB galaxies, since a 
larger fraction of the galaxy's light lies in the faint outer portions 
of the galaxies and larger errors in the photometry from these effects 
are inevitable.

The outermost isophote to which we can measure in each of our galaxies 
depends on a variety of factors: the intrinsic properties of the disk;
flatfield errors;
sky subtraction errors; observation in the presence of Moonlight or 
twilight; intrinsic variations 
in sky brightness due to the Milky Way and zodiacal light; and the presence of 
bright field stars.  We see no correlation between the color of the 
outer disk and the faintness of the last measured isophote as might be 
expected if sharper disk cutoffs are associated with younger stellar 
populations, while smoother edges are due to older 
stellar populations (Bosma 1983). 

We note 
in our sample that edge-on galaxies can typically be traced to fainter 
(face-on-corrected) surface brightness levels than less inclined
galaxies. However, to attempt to derive inclination-dependent 
magnitude corrections  based on this fact would be dangerous since
in a sample selected on {\it observed} rather than {\it 
intrinsic} surface brightness, the edge-on galaxies will always be 
among the intrinsically lowest (face-on corrected)
surface brightness systems, hence their nature may be somewhat 
different. Next, since sharp radial brightness cutoffs are 
known to exist for some brighter galaxies at a range of surface brightness 
levels (e.g., Barteldrees \& Dettmar 1994), 
it is possible that such cutoffs exist in at least some 
extreme late-type spirals as well. Moreover, the 
assumption of the symmetric, exponential nature of the disk in the outermost 
portions is highly questionable due to the difficulty in measuring 
these points in extreme late-type spirals. 
While it is true that for a given scale length, a galaxy with lower 
central surface brightness will contain a larger fraction of its light 
in the outer regions, pure extrapolations of fits to infinity can 
likewise introduce large errors for LSB disk magnitudes  in the 
absence of empirical justification (see Tully \etal\ 1996). Moreover, definitive
establishment of faint disk extensions of faint galaxies on small-field 
CCDs is extraordinarily difficult. Small amounts of scattered light, 
flatfield errors, and/or small errors in sky determinations can lead to 
grossly different characterizations of the outer disk profiles 
(e.g., Barteldrees \& Dettmar 1994).

Because our survey images are only moderately deep, in order to
empirically test these effects, we obtained deep V images of two of our 
sample galaxies (ESO~418-008 and ESO~305-009) 
using the CTIO Curtis Schmidt telescope in January 1995.
The Curtis Schmidt has a 
field of view of $\sim$25$'$. We obtained 6 exposures totalling 2500 
seconds of integration time for ESO~418-008 and 6 exposures totalling 
2400 seconds for ESO~305-009. 
We shifted the target to a different position on the chip between exposures. 
Flatfields on the Curtis Schmidt are excellent; the final frames have 
flatfields to better than 0.2\% and 0.5\% respectively.

We next compared the flux within an 
aperture identical to that used to photometer these galaxies 
in the 0.9m or 1.5m datasets. The result for ESO~418-008,
which had a limiting outer isophote of 26.6 mag~arcsec$^{-2}$ in the 
0.9m data, was that no additional flux was detected in the Schmidt data 
outside the adopted 0.9m aperture (Figure~3a). 
For ESO~305-009, additional flux was 
detected beyond the outermost 1.5m isophote (at 25.4 mag~arcsec$^{-2}$) 
only along about one third of the periphery of the isophote (Figure~3b). 
This light is not very extended; its detectable extent is 
$\sim$0.3$'$ on one side of the galaxy and $\sim$0.2$'$ 
on the other. This excess 
light accounts for only about 1\% of the total galaxy light. 

While these results cannot be strictly generalized to every galaxy in the 
sample, they do help to show that faint disk extensions are not likely 
to contribute a significant amount to the total luminosity of MLSB extreme 
late-type spirals in most cases. We contrast this with the LSB spiral 
sample of  
de~Blok \etal\ (1995), where
the difference between their measured aperture magnitude and their
magnitude derived from a radial light profile extrapolated to infinity 
ranged from 0.03 to 0.86 mags with a mean of 0.28 mags (an aperture 
correction of nearly 30\%).

\section{Discussion of Measured and Derived Optical Parameters}

\subsubsection{Colors and Color Gradients}
The derived aperture colors and total colors 
for our galaxies are presented in Table~5.
The spread in total \bmv\ (0.186--0.636) is consistent with 
that found in other samples of 
late-type spirals and irregulars with similar optical
luminosities (e.g., de Vaucouleurs {\it et al.}
1981; Gallagher \& Hunter 1986). We see no correlation of color with 
Hubble type (Figure~10). Our sample contains four extremely 
blue galaxies, (\bmv$<$0.3) which are interesting since these may be 
particularly young objects (cf. R\"onnback \& Bergvall 1994) and 
because galaxies with \bmv$\le$0.4  are too blue to 
be accounted for by stellar population models with constant star 
formation rates over time scales of $\ge$10~Gyr (e.g., Charlot \& 
Bruzual 1991; Kr\"uger \& Fritze-von Alvensleben 1994).  Colors 
show no correlation with inclination for most of
our sample (Figure~4) which further suggests that internal extinction 
is low. The exception is that two of the bluest galaxies in our sample 
actually have the {\it highest} derived inclinations. However, this is 
likely a selection effect since 
both galaxies have very low mean surface brightnesses and would be 
nearly invisible ($\bar\mu_{V}>$26.4 mag arcsec$^{-2}$) if viewed face-on.

Figure~5 shows a plot of color versus mean surface brightness for our 
sample. For most of the observed range in color, the two quantities show no 
correlation. Similar results have been found for field samples of 
late-type spirals and irregulars by de Vaucouleurs {\it et al.} 
(1981) and Gallagher \& Hunter (1986). Likewise, both sets of authors found 
the spread of color at a given Hubble type to be large as we do in our 
sample (see Table~5; Figure~10). The mean \bmv\ 
of our full sample (0.463$\pm$.118) is not 
significantly different from typical colors 
of higher surface brightness spirals of 
similar Hubble types in the literature. The existence of the two very 
blue galaxies (B--V$<$0.2) with very low mean surface brightnesses 
($\bar\mu_{V}>$26~mag arsec$^{-2}$) in our sample 
shows that not all of the very bluest galaxies are vigorously forming stars.

de Blok {\it et al.} (1995) also reported that their LSB spiral sample was not 
significantly different in terms of \bmv\ from HSB spirals, although 
they claim that \bmr\ colors showed their LSB samples to be bluer than 
HSB galaxies. This is surprising since  for
galaxies \bmv\ 
and \bmr\ are generally correlated. For example, the sample 
of R\"onnback \& Bergvall (1994), chosen to have the bluest \bmr\
colors 
in the ESO Catalogue, also showed extremely blue \bmv\ colors 
(mean \bmv$\sim$0.34).

From the colors in each of our three 
apertures, we can get an idea of the color gradient across the disk
(Table~5). We find 16
cases where the galaxy becomes {\it redder} with increasing r. One of 
the cases with outward reddening is a blue compact dwarf (BCD)-like object 
with a very blue, bright 
center ((B--V)$_{c}$=0.39) and faint, red outer arms ((B--V)$_{out}$=0.81). 
As is typical with BCD systems, 
the mean total color of this galaxy is fairly normal (\bmv=0.55) in spite 
of the outward reddening. Blue cores superimposed on
redder, lower surface brightness backgrounds appear to be
characteristic of some classes of starbursts (e.g., Hunter \etal\ 
1994). In 
the other 15 cases we can find no obvious galaxy traits which could 
explain the ``reverse'' color gradient and it is 
unlikely all of the cases can be attributed to extinction or 
inclination effects.  Objects with similar ``reverse'' color gradients
are also found in the 
sample of R\"onnback \& Bergvall (1994). Among spirals, this trend
appears to be a
unique feature of certain late-type disks (e.g., Tully {\it et al.} 
1996). Interestingly, this feature
is contrary to predictions of viscous disk evolution models, which 
predict radial bluing of the disk (e.g., Firmani {\it et al.} 1996).

\subsubsection{Surface Brightnesses}

We derived the the mean measured surface brightness for each galaxy from the 
formula:\\
\centerline{$\bar\mu_{\lambda}$ = m$_{\lambda} + $2.5~log($\pi ab$)}

\noindent where $a$ and $b$ are the semi-major and semi-minor axes of the 
galaxy in arcminutes, respectively, and m$_{\lambda}$ is the apparent 
B 
or V magnitude (see Table~3). 
We corrected this to a Galactic extinction-corrected
face-on value by assuming the galaxies are optically thin and applying 
the formula:\\
\centerline{$\bar\mu_{\lambda,face-on}$ = $\bar\mu_{\lambda}$ 
- 2.5~log(cos $i$) - A$_{\lambda}$}

\noindent where {\it i} is the galaxy inclination and  A$_{\lambda}$ is 
the Galactic extinction in the appropriate waveband (see Table~4).

The mean surface brightnesses of all of our sample galaxies span the 
ranges (in mag~arcsec$^{-2}$): $\bar\mu_{B}$=22.8--25.6 (measured); 
$\bar\mu_{B,f}$=22.9--27.0 (face-on- and Galactic extinction-corrected); 
$\bar\mu_{V}$=21.7--25.1 (measured); $\bar\mu_{V,f}$=22.4--26.8 
(face-on- and Galactic extinction-corrected).

\subsubsection{Disk Scales}
Because we did not compute azimuthally averaged light profiles for our 
sample galaxies, we have not derived traditional exponential 
scale lengths. 
Nonetheless, in order to provide some measure of the steepness and
extent of the light profiles of the galaxies, we have 
derived a ``mean disk scale'' $\bar\alpha_{V}$ defined as: 
($\bar\mu_{V1} - \bar\mu_{V2}$)/1.086r, 
where $\bar\mu_{V1}$
is the mean surface brightness in V between the middle and the 
outermost fitted ellipses at r$_{1}$, $\bar\mu_{V2}$ is the mean V surface
brightness within the innermost fitted ellipse at r$_{2}$, 
and r = r$_{1}$ -- r$_{2}$.
This measure reduces to a normal scale length in the case of a pure 
exponential disk. Our derived mean disk scales are given in Table~4. On 
average these values are smaller than the exponential scale length  in 
other recent LSB spiral samples (e.g., McGaugh \& Bothun 1994;
de~Blok \etal\ 1995), suggesting 
that our respective samples contain somewhat different (i.e., 
physically smaller) types of 
galaxies.

\subsubsection{Correlations between Optical and HI Properties}
In order to gain further clues on the evolutionary status of our sample 
objects, we have made a comparison of their optical and \HI\ 
properties.
Figure~6 show a plot of \bmv\ versus \dark, the ratio of the 
neutral 
hydrogen mass to the  optical luminosity. 
It can be seen that the 
bluest galaxies in the sample all have relatively high neutral gas fractions, 
consistent with 
their status as young objects, although high \dark\ 
galaxies are found over the full range in B--V color represented by our 
sample. The lack of blue galaxies with low \dark\ is likely to be a 
real effect since originally all targets were selected from 
{\it blue}-sensitive photographic plates.
The degree of scatter in the \dark\ values among the redder sample 
galaxies (two orders of magnitude at \bmv$\sim$0.6) is surprising since 
it indicates that galaxies whose colors are dominated by older stellar 
populations can still retain large gas fractions. These might be 
galaxies with large gas reservoirs in their outer disks where star 
formation cannot occur efficiently (e.g., Hunter \& Gallagher 1986). This
suggests that either the extreme late-type spirals represent a wide range 
in star-formation histories or that gas and star formation in extreme 
late-type spirals cycle in a manner different from current models.

Several of our sample galaxies have extremely high \dark\
ratios ($\sim$3-9 in solar units); such values are reminiscent of
that of the ``\HI\ Cloud'' 
in Virgo (\dark$\sim$10), which is believed to be a very young object 
with
an age of only roughly 1~Gyr (Salzer {\it et al.} 1991). This further 
hints that 
some of our sample galaxies may be extremely young. For the bulk of the 
sample however, the \dark\ ratios are reminiscent of field samples of 
irregulars with moderately young stellar populations
(e.g., Gallagher \& Hunter 1986).

Further insight comes from examining \dark\ versus mean 
inclination-corrected
surface brightness ($\bar\mu_{V,i}$) (Figure~7).  This plot shows that
\dark\ and surface brightness are correlated for
galaxies spanning a wide range of \bmv\ color.
Thus while the highest 
\dark\ galaxies in our sample may represent different evolutionary 
histories, or different stages in the stellar evolution in a single 
family of objects, they all share the property of being some of the 
lowest surface brightness and structurally diffuse galaxies in our 
sample. Similar trends have also been found
in samples of irregular galaxies (see Gallagher \& Hunter 1985).

\subsection{General Trends in the Sample}
\subsubsection{Offset Centers}
One of the features which distinguishes
extreme late-type spirals from irregulars is that their optical centers
can be unambiguously located through the presence of a nucleus or a 
nuclear region of enhanced brightness. However, in  about 
two-thirds of our sample this light ``center''
is offset from the center of the outermost isophote of the galaxy by 
anywhere from 2-15$''$, indicating that intrinsic 
optical asymmetries occur frequently in extreme late-type spirals. Such 
asymmetries have been previously noted to be commonplace in other Magellanic 
spirals (e.g., Odewahn 1996) and extreme 
late-type Scd-Sdm spirals (e.g., Karachentsev {\it et al.} 
1993).

An interesting question which remains to be explored is 
whether the light center is the true dynamical center of the galaxy. 
Matthews {\it et al.} (1997b) investigate how these optical asymmetries 
compare with the asymmetries seen in many of our sample galaxies in 
high-resolution, high signal-to-noise, single-dish \HI\ spectra (see 
also Richter \& Sancisi 1994).

\subsubsection{Bars}
An examination of Plates 1-7 reveals that
unlike in the LSB spiral sample of McGaugh {\it et al.} (1995), bars are not 
rare in our sample (see also Table~4);
37\% of our sample (18 galaxies) is barred, and we have classified an 
addition 11 objects as transition objects (indicated $AB$ in Table~4). 
In some cases the bars in our objects 
are slightly off-center, like those common in Magellanic
irregular galaxies. In other cases, the bar is fairly centrally
located as is typical in spirals. 

We find that the occurrence of bars shows no 
correlation with mean disk surface brightness, contrary to the 
theoretical arguments of Mihos {\it et al.} (1997) who show, based on 
surface density, that LSB spirals should be very stable against the 
formation of bars. On the other hand, given the high frequency of 
bars  in irregulars (e.g., de~Vaucouleurs \& Freeman 1972), 
and in Sd-Sm spirals (e.g., Feitzinger 1980; 
Odewahn 1996) empirically it is not surprising that they should be common 
among extreme late-type spirals.  Such a trend was also noted by 
van~den~Bergh (1966) for DDO galaxies; bars are more frequent among 
small, faint DDO spirals than among larger, brighter DDO spirals. 
If  extreme late-type spirals have very large dark halo 
cores compared to their disk scale lengths (as 
rotation curve measurements of similar galaxies suggest) then 
the absence of shear in the inner region could facilitate bar formation
and increase a bar's longevity (Sparke 1997).

Our observation may also reflect the fact that barred objects 
tend to be excluded from central surface brightness-selected samples
such as that of McGaugh {\it et al.} (1995).
Dynamical studies are needed to 
further explore the nature of the bars in extreme late-type spirals.

\subsubsection{Surface Brightness ``Steps''}
Bosma \& Freeman (1993) noted the existence of multi-component
disks among LSB spirals. We see numerous examples of this
phenomenon in our data (see Appendix~2; Plates 1-7). Because 
the greyscale images shown on the Plates are stretched to best 
emphasize detail in the central regions of the galaxy, the
``steps'' are more difficult to discern. We therefore show in 
Figure~8 one example, ESO~358-020, with a greyscale 
stretch that emphasizes the ``stepped'' light profile. 

The existence of these surface brightness ``steps''
means that the surface brightness profiles of the galaxy cannot be 
adequately fit by a single exponential. This is best illustrated
in a major axis profile plot rather than an azimuthally averaged radial 
profile
since if the different disk components are not perfectly concentrically 
elliptical isophotes, 
radially averaging tends to smear out these gradients (see Figure~9).

The existence of surface brightness steps and general deviations from 
single exponential behavior of the disks, may imply a unique evolutionary
status of these disks; in particular, it may be an additional signature 
of youth. In many giant spiral galaxies, the optical radial surface 
brightness profiles are close to exponential over at least 3-4 radial 
scale lengths (e.g., de Jong 1996). This implies that some mechanism 
has operated in the disks of giant galaxies to produce approximately 
exponential stellar density surface brightness profiles, while within 
the same disks the \HI\ often shows a much slower decline in density 
with increasing radius. An explanation for this phenomenon by Lin \& 
Pringle (1987) is that exponential disks result when the star formation 
and viscous times scales are comparable (see also Struck-Marcell 1991; 
Firmani \etal\ 1996). The presence of multi-component disks in the 
small spirals studied here then indicates that some type of break-down 
has occurred in this process.

One possible way to eliminate the single exponential disk might be to 
have a very low effective viscosity. In this case the gas would remain 
in a distribution determined by its initial angular momentum. 
Presumably this material could also form stars, albeit at low rates, 
and thereby possibly produce a flat outer disk.  Galaxies with 
disk ``steps'' might therefore be considered to be dynamically unevolved. 

The present data hint that color gradients are associated with the 
surface brightness ``steps'', although our sample is too small to 
establish 
whether these gradients are typically steeper or more abrupt than in galaxies 
without the ``step'' features. However, there is some indication that 
the ``stepped'' galaxies more often show reverse color 
gradients (in the sense of getting redder rather than bluer at larger 
radii; Section~5.0.1). Eight out of 14 ``stepped'' galaxies get 
redder with increasing r (57\%), 
while only 8 out of 29 (28\%) of the galaxies without a ``stepped'' surface 
brightness distribution exhibit radial reddening. Perhaps then 
the faintest outer disk component in these galaxies may be
predominantly composed of older stellar populations.

\subsubsection{Point-like Nuclei}
Another trend which emphasizes the true spiral-like nature of our sample 
objects is that 
unresolved, point-like nuclei are relatively common.
van den Bergh (1995) noted that 
the existence of such nuclei may be a distinguishing characteristic 
between spiral and irregular galaxies. Ten 
of our forty-nine objects in our sample have such 
nuclei, and five additional cases we list 
as possible nuclei candidates (see Appendix~2). From H$\alpha$--R-band images 
of most of our nucleated galaxies 
(Matthews \& Gallagher, unpublished) we are able to ascertain that the 
central unresolved objects are not unfortunately placed foreground stars.
Moreover, that we can even see 
these faint nuclei suggest the host galaxies 
have little or no bulge component, (as also is the case in the M33-type 
Sc~III spirals) and that the internal extinction, even within the 
central regions of the galaxies, is very low (see also Matthews \etal\ 
1996; Gallagher \etal\ 1997). 

The existence of compact nuclei in extreme late-type spirals is 
surprising, since compact nuclear concentrations are thought to require 
strong central gravitation potentials, seemingly inconsistent with the 
optically diffuse centers of extreme late-type spirals. The nuclei may 
therefore indicate the presence of centrally concentrated dark matter.
Additional spectroscopic observations and imaging with space-based
resolutions  are needed to establish the true nature of these nuclei.
It is possible that in some cases the nuclei may be very low-luminosity
AGNs like that in the nearby Sd~IV spiral NGC~4395 (e.g., Filippenko 
\& Sargent 1989; Matthews 
{\it et al.} 1996; Gallagher \etal\ 1997) or M33-like compact 
starburst nuclei (Kormendy \& 
McClure 1993). An example of the latter is seen in the nearby extreme 
late-type
Sdm~IV spiral NGC~4242  (Matthews {\it et al.} 1996; Gallagher \etal\ 
1997).

\subsubsection{Looking for Evidence of Bulges}
It is well-know that the bulge component becomes decreasingly 
important among the latest spiral types. Eye inspection of our images 
reveals only five candidates for bulges (see Appendix~2); 
moreover, all but one 
of these (ESO 440-049) are questionable calls. Justification of 
true bulges in MLSB galaxies in the absence of kinematic data can be 
deceptively difficult, especially for nearly face-on galaxies.

In face-on systems, it is difficult to distinguish bulges from 
multi-component disks (see Section~5.1.3), although a circularization of 
the isophotes near the center is one signature of a bulge. 
Schombert {\it et al.} (1995) 
claim reddening in the central regions of their ``dwarf spirals'' as evidence 
for bulges, but this claim is not substantiated by their Figure~5. 
Moreover, the degree of reddening in the central portions of a galaxy 
correlates very poorly with bulge size (cf. de~Jong 1995), so this is 
not a definitive test.

We conclude that among the extreme late-type spirals, bulges are most 
often very weak or non-existent although they are more common in 
larger LSB disks, generally of earlier Hubble type and larger scale 
lengths (e.g., Knezek 1993; Sprayberry {\it et al.} 1995b). Eventually 
stellar velocity dispersion measurements are needed to better explore this 
issue.

\section{Are There Multiple ``Families'' of Faint Disk Galaxies?}
The giant, very low surface brightness
``Malin~1''-type objects are among the most luminous spirals known
(e.g., Sprayberry {\it et al.} 1993). It seems clear that 
these Malin-1-type objects
form a distinct family of LSB galaxies with a unique formation history (e.g., 
Hoffman {\it et al.} 1992) and no 
HSB analogues (e.g., McGaugh {\it et al.} 1995).

Knezek (1993) studied a sample of large LSB spirals, the majority of 
which would not be considered ``Malin-1''-type objects, but which were 
nonetheless selected for their high \HI\ masses  and large scale length 
disks. These spirals have luminosities comparable to HSB giants, and 
often possess significant bulge components.
Many of the LSB spirals in other recent samples (e.g., McGaugh \& 
Bothun 1995; de Blok \etal\ 1995) could 
also be categorized with these objects (hereafter LSB 
giants).  Current results suggest that 
the rotation curves of 
LSB giants are similar to radially 
scaled rotation curves of giant HSB galaxies 
(e.g., de Blok \& McGaugh 1997; Salucci \& Persic 1997).
 
How are  these LSB giants related to the extreme late-type 
spirals we present here? At first glance, it may seem that the natural 
assumption is that these are all members of a single family of 
objects with the extreme late-type spirals being simply 
scaled-down versions of LSB giants. However, 
Wirth \& Gallagher (1984) and Kormendy (1985) showed 
how this type of logic was misleading for
lowest luminosity elliptical galaxies. They found that there 
exist two distinct 
families of small ellipticals, one similar to scaled-down versions of 
the giant ellipticals, and the other more 
structurally diffuse, and having distinct photometric properties.

At present there are several hints that there are structural
differences between LSB giants and the extreme late-type spirals. One 
clue is the trend in the Tully-Fisher (TF) relation 
for extreme late-type spirals found by Matthews {\it et al.} 
(1997a). Sprayberry {\it et al.} 
(1995a) and Zwaan {\it et al.} (1995) both explored the TF relation for 
samples of LSB spirals and found them to be indistinguishable from that 
of HSB spirals (see also de~Blok {\it et  al.} 1996). However, Matthews 
{\it et al.} (1997a) found the extreme late-type spirals (i.e., the 
forty-nine
galaxies presented here, plus five additional objects from R\"onnback \& 
Bergvall 1994) to fall, on average, 1.3 magnitudes below the TF 
relation defined by HSB galaxies.  Matthews {\it et al.} (1997a) argue 
that this difference results primarily from the selection of two 
different galaxy populations: samples of predominantly LSB giants 
versus the extreme late-type spirals.

It appears then that the tendency 
to deviate from the TF relation (in the sense of a galaxy being 
underluminous for its rotational velocity) increases markedly among the 
smallest and least luminous MLSB disks---i.e., among the extreme 
late-type spirals. Further evidence for this trend comes from the fact 
that in our sample TF deviation is strongly correlated with mean disk 
scale and with optical size in the sense that the 
smallest galaxies are the greatest 
deviators and that there exist weak 
trends of TF deviation increasing with decreasing \HI\ mass and with 
increasing \dark\ ratios (Matthews {\it et al.} 1997b).

The latter trends suggest an increasing predominance of 
dark matter in these systems. This finding is consistent with rotation curve 
analyses and mass modelling which exists for a few examples of extreme 
late-type galaxies in the literature (e.g., Carignan 1985; Jobin \& 
Carignan 1990; C\^ot\'e 
{\it et al.} 1991; Broeils 1992; Martimbeau 1994; Rownd {\it et al.} 1994) 
and with the two TF deviators (DDO~154 and NGC~2915)
discussed by Meurer {\it et al.} (1996). 
Meurer {\it et al.} (1996) further argue that the 
the BCD/late-type spiral NGC~2915 would never fall on 
the TF relation, even if all of its gas were converted {\it in situ} to 
stars, implying a structural uniqueness in this type of galaxy.
Indeed, studies of  the smallest 
spiral galaxies may be the key to furthering our understanding of dark 
matter in galaxies (e.g., Kormendy 1987; Salucci \& Persic 1997).

In summary, current evidence suggests that {\it extreme 
late-type spirals are not simply scaled-down giants.}
Ultimately, further detailed dynamical  studies will help make to
determine how different giant LSBs and extreme late-type spirals 
are in terms of their structure and mass distribution. 

\section{Connections with the More Distant Universe}
\subsection{Possible Links between Extreme Late-Type Spirals and Faint 
Blue Galaxies}
About half of the galaxies in our sample have 
\bmv$\leq$ 0.5, making them bluer than the typical lower bound
to colors of giant spirals (de Jong 1995). This raises the 
possibility that some of the galaxies in our sample could be similar to the 
redshift z$>$0.3 ``faint blue galaxies''.

Whether small,
extreme late-type galaxies comprise a significant fraction of the
faint blue galaxies depends on the form of the luminosity function
and on evolution of lower luminosity galaxies with lookback time,
neither of which is currently
well-determined (e.g., Marzke \etal\ 1994b; Heyl \etal\ 1997; Lin \etal\
1997).  Since we do not have a
quantitative measure of the local space densities of extreme late-type 
spirals, we limit the present discussion to a
preliminary exploration of the potential visibility of typical members 
of our sample at moderate redshifts. A full treatment of this topic
requires evolutionary models and an evaluation of the 
redshift-dependent
k-correction.

Consider as an example the
observable properties of typical extreme late-type spirals seen at
a redshift of z$\approx$0.8 in a no-evolution model.
At this redshift the B-band in the rest
frame shifts to the I-band in the observer's frame. Adopting standard 
calibrations for B and Kron-Cousins I band (see
Bessell 1993), we find that the observed and intrinsic mean surface
brightness are related by
$$ \mu_I(observed) \approx \mu_B(rest) +0.7.$$
Thus a blue, intermediate surface brightness Sd-Sm galaxy
from our sample, located at z$=$0.8, would be observed to have
$\mu_I =$23--25
mag~arcsec$^2$ over a central region with a diameter of 2--5 kpc.  For 
a low density Universe with q$_0 =$0.05
this region would have an angular sizes
of about 0.3--0.8$''$, or well above the resolution limit of
WFPC2 on the {\it HST}.

The isophotal luminosities  of the inner, higher surface brightness
central regions of some of the brighter extreme late-type spirals in 
the present sample have M$_B \approx$ -16 -- -17, 
corresponding to I$\approx$27 at z=0.8.
If galaxies of this type are present at redshifts z$<$0.8,
then they would be detectable
in deep WFPC2 {\it HST} exposures,
such as the Hubble Deep Field (Williams {\it et al.} 1996).
Also, the small sizes and moderate-to-low surface brightnesses predicted
for the cores of such galaxies are consistent with the properties
of many of the faint galaxies detected in deep 
WFPC2 exposures by Odewahn {\it et al.} (1996).  For comparison,
analogs to common, nearby  dwarf irregular galaxies will tend to be
fainter when observed at moderate redshifts as a result of their 
generally
low central surface brightnesses and lack of central light 
concentration.

We conclude that galaxies at z$=$0.8 with similar properties to present-day
Sd-Sm systems would be observable with WFPC2 on {\it HST}. These galaxies
would become increasingly detectable at lower redshifts, where they 
would
be blue objects with moderately small angular sizes.
A further and more detailed analysis that is beyond the scope
of this paper is required to see if this or a related class of galaxy
could be sufficiently frequent as to be a substantial component of the
faint blue galaxy populations.

\subsection{Lessons on Morphological Classification at Higher 
Redshifts}
The large range in physical sizes and surface brightnesses of extreme
late-type galaxies means that distances to these objects are difficult
to estimate from optical images. Many extreme late-type galaxies that
are relatively nearby can easily be mistaken for distant giant
systems of earlier structural types (e.g., ESO 418-008 could be 
misclassified as a distant giant SBc spiral).  This effect
can be important, for example, on Schmidt survey plates
where many galaxy images have low signal-to-noise ratios
or very small angular sizes, and thus are 
difficult to classify. Such an effect may account in part for different
shape of the local galactic luminosity functions found
in different  surveys,  since morphological confusion adds to the
difficulties in making luminosity functions in which galaxies
are sorted by structural class (e.g., Bingelli \etal\ 
1988; Marzke \etal\ 1994a).

The large scatter in relationships between morphology and other 
properties
of extreme late-type galaxies may also have implications for the
interpretation of populations of faint blue galaxies. A considerable
effort has gone into determining the structures of faint galaxies
found in various deep exposures with the WFPC2 on the {\it HST} 
(e.g., Driver \etal\ 1995; Abraham \etal\ 1996;
Odewahn \etal\ 1996).  While these data are proving very valuable
as a means to understand {\it populations} of galaxies at moderate-to-high 
redshifts, our results suggest they may be of limited utility
in defining the {\it properties} of individual galaxies.

\section{Discussion and Summary}
We have presented new B and V aperture photometry for forty-nine small, 
late-type, moderate-to-low surface brightness field spirals. We have 
termed these galaxies ``extreme late-type spirals''.
Unlike other recent photometric 
studies of LSB disk galaxies, this study focussed 
on a morphologically selected sample. Our goal was to gain insight
into the nature of some of the smallest and least evolved spiral disk 
galaxies. These are not a new class of galaxy (see, e.g., Briggs 1997) 
but recent \HI\ and optical surveys have revealed their past 
under-representation in Local Supercluster field samples.
Data 
such as we present here are a 
necessary first step in the selection of samples for more detailed 
studies of individual properties, such as \HI\ synthesis observations, 
optical spectroscopy, and mass modelling.

We find that extreme late-type spirals do not appear to be simply 
scaled-down versions of giants, but rather they are
in many ways intermediate between irregulars and
HSB giant spirals of similar Hubble types in terms of their global 
properties. In spite of their 
differing degrees of disk organization, both extreme 
late-type spirals and irregulars (as well as other MLSB spirals such as 
the DDO dwarfs)
span similar ranges in optical luminosity, optical surface
brightness, \HI\ content, velocity width, and \bmv\ colors (cf., 
Figure~10; van den Bergh 1966; de Vaucouleurs \etal\ 1981; Longmore 
\etal\ 1982; Gallagher \& Hunter 1986).

In terms of disk structures, extreme late-type spirals are diverse;
some have very faint 
but clearly defined arms, others have only rudimentary arms; 
others have very diffuse, 
structureless disks. This trend is 
obvious from Plates 1-7 and Table~4. Types of objects include
spiral/irregular transition objects (e.g., 
ESO~430-039), compact, BCD/spiral transition objects (e.g., 
ESO~504-017), thick Magellanic spirals 
(e.g., ESO~418-009), and thin, extended Sd systems which appear to have
dynamically very cold disks (e.g., ESO~482-005). For the bulk of our 
sample, there is no obvious link 
between the optical morphology and the
derived optical or \HI\ parameters (e.g., Figure~10). One 
of the unsolved questions is why there exists this 
diverse range of morphologies in an otherwise similar physical parameter space 
(see also Gavazzi \etal\ 1996). 

\HII\ region 
spectroscopy, which exists in the literature for a few analogous objects 
(e.g., Goad \& Roberts 1981; R\"onnback \& Bergvall 1995), 
suggests that extreme late-type spirals also generally have very 
low metallicities, comparable to irregulars.
Rotation curve analyses and mass modelling
(presently available only for a few objects meeting our definition of 
extreme late-type spirals) suggest that like 
irregulars, extreme late-type spirals are often dark matter dominated, 
even within the optical galaxy (e.g., Carignan 1985; Broeils 1992; 
Meurer \etal\ 1996).
One piece of indirect 
evidence that this trend is widespread among extreme late-type 
spirals is that  many of these galaxies fall below the 
TF relation for brighter galaxies (e.g., Matthews {\it et 
al.} 1997a), consistent with other dark matter-dominated late-type 
spirals and irregulars (e.g., Meurer {\it et al.} 1996). Other hints 
of dark matter domination are that
these galaxies can harbor compact nuclei, the apparent longevity of 
bars, and that these 
structurally diffuse disks can maintain organized structures over a 
Hubble time. 
Together the aforementioned findings suggest that extreme late-type 
spirals include some of the least evolved spiral galaxies in terms of 
both star formation and dynamical properties.

From the derived properties of the present data we note several 
other interesting trends among extreme late-type spirals
galaxies: (1)  bars are common features 
[18 out of 49 galaxies in our 
sample are barred; 11 more objects are intermediate ($AB$) galaxies];
(2) unresolved nuclei are found in 10 of our galaxies; these may 
be related to low-level AGNs or M33-like compact starburst nuclei;
(3) the observed range of \bmv\ colors for our 
sample is similar to that of high surface brightness and moderate 
surface brightness samples of similar Hubble types; color is not 
well-correlated with surface brightness; we have four very blue objects in 
our sample (\bmv$<$0.3)
which may be extraordinary young objects; (4) a number of our galaxies 
have large \HI\ fractions (\dark$>$3); these objects include the three 
bluest galaxies in our sample, but also include galaxies spanning the 
full range of observed \bmv\ colors found in our sample; large \dark\ 
also seems to correlate with low surface brightness over a range in 
color; (5) several of our 
sample galaxies show disk reddening with increasing radius; (6) many of our 
disks cannot be well fit with a single exponential disk even though 
they do not possess a bulge component; (7) sharp 
surface brightness gradients (``steps'') are seen in a number of our 
galaxies; (8) the light centers of extreme late-type spirals are often 
displaced from the outermost galaxy isophote; (9) extreme 
late-type spirals appear to be a structurally unique family of galaxy 
rather than simply scale-down giant spirals;  
(10) higher-redshift analogues of
moderate surface brightness extreme late-type spirals 
may comprise at least part of the faint blue galaxy populations.

\acknowledgements
We thank CTIO for their generous allocation of observing time for this 
project. We also thank D. M. Peterson for a critical reading of this 
manuscript. This research was funded by the Wide
Field and Planetary Camera 2 Investigation Definition Team, which is
supported at the Jet Propulsion Laboratory (JPL) via the National
Aeronautics and Space Administration (NASA) under contract No.
NAS7-1260 as part of a project to identify new classes of targets for 
WFPC2 observations.

\appendix
\subsection{Appendix: Error Estimation}
For each galaxy we have computed an error for our derived magnitude 
(columns 3 \& 5, Table~3)
which consists of the sum in quadrature of the errors resulting from 
(1) shot noise, (2) read noise, (3) small-scale pixel-to-pixel variations 
in excess of Poisson noise, and (4) flatfield errors.  Our 
estimates refer to internal errors only, and do not take into account 
uncertainties in the photometric solution (see Table~2). An additional 
discussion of error analysis for galaxy CCD photometry may be found in 
Knezek (1993).

For 
some galaxies, additional uncertainty occurs if stars are present in 
the galaxy aperture (since they cannot be removed perfectly) and/or if 
the galaxy is large compared with the field-of-view of the CCD (and 
therefore sky flux cannot be accurately measured on the images without 
contamination from the galaxy light). Both of these errors are 
difficult to quantify, so we simply flag these cases in Table~3.

We have computed our magnitude errors from the following general formula:
\centerline{(1)~~~~~~~~$\sigma_{mag} = -2.5~log_{10}(\frac{N_{SOURCE}}
{N_{SOURCE}\pm N_{NOISE}})$}

\noindent where N$_{SOURCE}$ is the galaxy flux within the chosen aperture and 
N$_{NOISE}$ is the sum of all all noise sources (in electrons) within 
the aperture. For this work we assume that:

\centerline{N$_{NOISE} = \sqrt{\sigma^{2}_{POISS} + \sigma_{SS}^{2} + 
(\sigma_{FF})_{Sky}^{2} + (\sigma_{FF})_{SOURCE}^{2}}$}

\noindent We discuss each of these terms below.

\subsubsection{Poisson Noise}
\noindent In general for CCD observations, 
\centerline{$\sigma_{POISS} = \sqrt{total~electrons~counted} = 
\sqrt{\frac{N_{SOURCE}}{G} + \frac{N_{SKY}}{G} + (n_{pix}\cdot 
RN^{2})}$.}

\noindent where N$_{SOURCE}$ is the total number of electrons 
in the source aperture, G 
is the CCD gain in electrons per ADU, N$_{SKY}$ is the number of sky 
electrons in the aperture, measured within several ``sky boxes'' 
as described in Section~3.4.2,
n$_{pix}$ is the number of pixels in the galaxy aperture, and RN is 
the CCD readnoise in RMS electrons per pixel.

\subsubsection{Small-Scale Pixel-to-Pixel Noise}
We measured small-scale pixel-to-pixel noise in excess of Poisson noise 
by computing the RMS value for all of the pixels 
within each individual sky box measurement ($\sigma_{BOX}$)
and comparing this with the error expected from pure Poisson statistics 
($\sigma_{BOX,POISS}$). 
Thus:

\centerline{$\sigma_{SS} = \sqrt{\sigma_{BOX}^{2} - 
\sigma_{BOX,POISS}^{2}}$}

\noindent where $\sigma_{BOX}^{2}$ is taken to be the sky box with the highest 
measured internal standard deviation each particular image. For most galaxies 
this term was negligible, unless there were many field stars or faint 
background galaxies in the frame.

\subsubsection{Flatfield Errors}
Because flatfield errors affect both the sky count measurement and the 
object count measurement, flatfield errors consist of two components.
We estimated the effect of flatfield errors on our sky measures by 
comparing the standard deviation between the mean sky value determined 
from  the different sky box measurements on each image
with the deviation expected from pure Poisson statistics. The standard 
deviation between sky boxes is expressed as:

\centerline{$\sigma_{BBOX} = \sqrt{\frac{(\bar N_{box 1} - \bar 
N_{B})^{2} + (\bar N_{box 2} - \bar N_{B})^{2} + ...}{N_{tot} - 1}}$}

\noindent where N$_{tot}$ is the total number of sky boxes measured, $\bar 
N_{box,n}$ is the mean counts per pixel in box n, and $\bar N_{B} = 
\frac{N_{box 1} + N_{box 2} + ...}{N_{tot}}$.

Assuming the distribution  of pixel values within the sky box is
Poisson, then each sky box pixel can be treated as an independent sky measure 
and we have

\centerline{$\sigma_{BOX,POISS} = 
\frac{\sqrt{total~sky~counts~in~sky~box}}{n_{pix,sb}}$.}

\noindent Here, n$_{pix,sb}$ is the number of pixels within the sky 
box.

Over the entire galaxy aperture, the noise due to the flatfield 
uncertainty on the sky measurement is given by the difference, in 
quadrature, of the above two terms times $n_{pix}$, where n$_{pix}$ is 
the number of galaxy aperture pixels.

\centerline{$(\sigma_{FF})_{SKY} = [\sqrt{\sigma_{BBOX}^{2} - 
(\sigma_{BOX})_{POISS}^{2}}] \times n_{pix}$}

The effect of the flatfield errors on the object flux measurement can 
be estimated simply by scaling $(\sigma_{FF})_{SKY}$ by the ratio the source 
counts ($N_{SOURCE}$) to the sky counts within the galaxy aperture 
($N_{SKY}$):

\centerline{$(\sigma_{FF})_{SOURCE} = (\frac{N_{SOURCE}}{N_{SKY}})
\cdot (\sigma_{FF})_{SKY}$.}

\subsubsection{Final Formula}
Finally, combining all noise terms by summing in quadrature, 
we arrive at the formula

\centerline{N$_{NOISE} = \{\sigma_{POISS}^{2} + \sigma_{SS}^{2} + 
n_{pix}^{2}(\sigma_{FF})_{SKY}^{2}(1 + 
(\frac{N_{SOURCE}}{N_{SKY}})^{2})\}^{\frac{1}{2}}$.}

\noindent which may be used in Equation (1).

\subsection{Appendix: Comments on Individual Objects}
\begin{flushleft}
{\it ESO 547-020:} very flocculent structure; weak
bar.

{\it ESO 418-008:} barred; moderately high surface
   brightness but very tiny spiral.

{\it ESO 418-009:} bright bar traced by a
fairly HSB region; disk very flocculent and very symmetric.

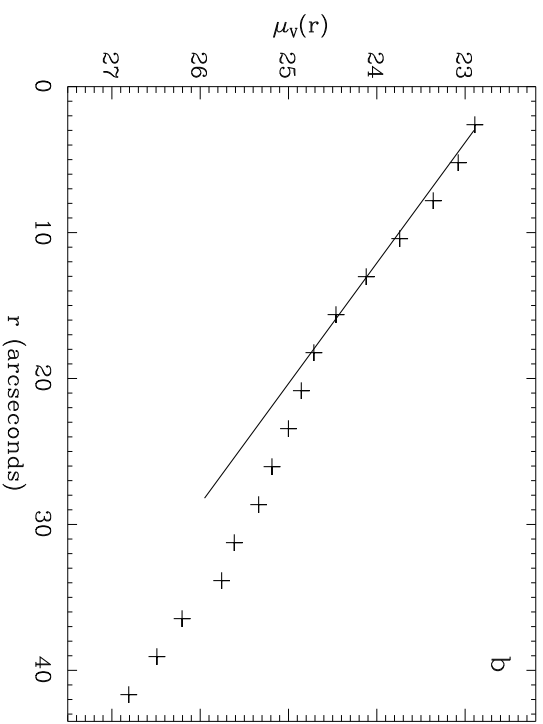
{\it ESO 358-020:} point-like nucleus; barred; very
twisted light distribution;
HSB center with three moderately bright, compact sources embedded;
2-step disk.

{\it ESO 549-002:} appears to contain many \HII\
regions; 3-step disk with
irregular edges; very smooth underlying light distribution.

{\it ESO 359-016:} edge-on; 3-step disk;
possibly barred; asymmetric light distribution.

{\it ESO 359-031:} barred; 2-step disk.

{\it SGC 0448-395:} point-like nucleus; 2-step disk;
outer disk extraordinarily
faint with no discernible structure.

{\it ESO 422-005:} =T0450-28; 2-step, diffuse disk.

{\it ESO 552-066:} point-like nucleus;  very faint
diffuse disk with a very slight central light enhancement.

{\it NGC~2131:} =SGC 0556-263;  peculiar; very bright
center with
peculiarly-shaped, very diffuse disk; BCD candidate.

{\it ESO 425-008:} =SGC 0604-275; very faint, barred
edge-on.

{\it AM 0605-341:} 4-step, structureless disk with
a very bright bar; BCD candidate.

{\it ESO 497-007:} =AM 0902-253; point-like nucleus;
very faint; possible arm structure visible.

{\it ESO 373-026:} bright bar; well-defined, fairly
loosely-wound arms.

{\it ESO 500-032:} light from a very bright star
occults part of galaxy; point-like nucleus set in brighter central 
region; very
faint, diffuse disk with no discernible arm structure.

{\it ESO 377-017:} =AM 1107-363; nucleus? very
structureless disk; little central light concentration.

{\it ESO 503-022:} nucleus? peculiar; 2-step disk;
bright bar traced by
a moderate surface brightness region with an extremely faint disk
extension; no discernible arms.

{\it ESO 504-010:} barred; flared and/or warped disk.

{\it ESO 504-017:} =AM 1146-270; very bright center
surrounded by faint,
tightly-wound arms; likely a BCD or related object.

{\it ESO 440-039:} point-like nucleus; asymmetric
light distribution; weak
bar; somewhat irregular, lopsided spiral.

{\it ESO 440-049:} nearly face-on; dE-like companion 
present
with m$_{V}$=18.0; system
resembles LSB analogue of M51; likely has a bulge component;
outer disk
is very faint and symmetric; clearly visible arms.

{\it ESO 441-023:} =AM 1213-311; very faint, irregular
arms; 3-step disk.

{\it ESO 380-025:} two-piece bar-like feature; disk
appears flared; extent
of disk is symmetric but light distribution is very asymmetric.

{\it ESO 443-080:} one-armed spiral with a
well-defined single arm;
has somewhat brighter nuclear region containing what appear to be two
bright \HII\ regions.

{\it ESO 508-034:} very bright center; possibly
contains a bulge.

{\it ESO 445-007:} extraordinarily faint disk.

{\it ESO 510-026:} point-like nucleus; very faint;
arms visible only in V frame;
possible weak bar.
                                                                
{\it ESO 446-053:} symmetric disk but with very
asymmetric light distribution;
appears to have many small \HII\ regions.

{\it ESO 482-005:} near edge-on; barred; 2-step disk;
outer disk very faint;
possible dwarf companion.

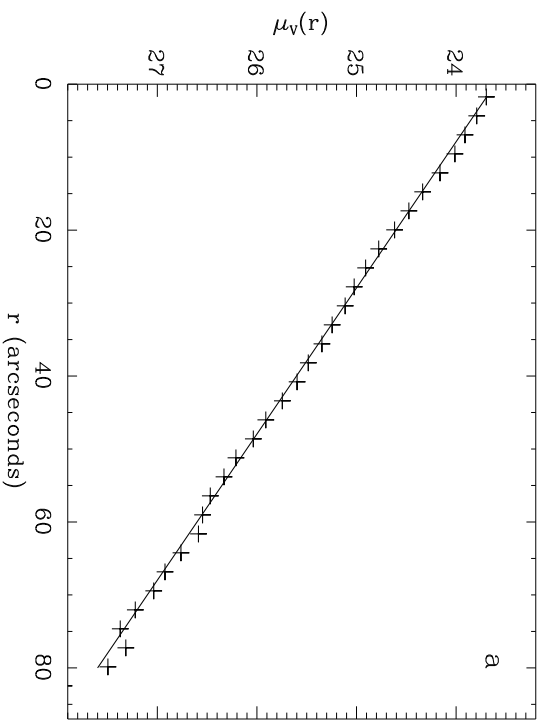
{\it ESO 358-015:} point-like nucleus; very tiny 
2-step disk
with asymmetric light distribution; bulge?

{\it ESO 548-050:} nucleus? near edge-on, but arms
visible; fairly HSB
center; asymmetric light distribution.

{\it ESO 358-060:} diffuse; nearly edge-on; fairly
uniform surface brightness;
possibly an irregular.

{\it ESO 359-029:} point-like nucleus; 2-step disk;
small, very
smooth, diffuse disk.

{\it ESO 305-009:} =T0506-38; point-like nucleus;
fairly diffuse disk
with one faint arm; possible bulge or bar.

{\it ESO 487-019:} barred; arm structure superposed on 
smooth
disk.

{\it ESO 488-049:} =SGC 0556-252; barred; very faint
outer disk;
appears to contain many small \HII\ regions throughout disk.

{\it ESO 431-015:} very high Galactic extinction;
little central light
concentration;
inclination and morphological classification uncertain.

{\it ESO 318-024} =AM 1055-391; peculiar ``X'' shape;
possible merger remnant;
very large bar or elongated disk structure
surrounded by a diffuse outer disk or halo; inclination very
uncertain.

{\it ESO 502-016:} barred; small, diffuse disk.

{\it ESO 438-005:} =T1106-28; very faint, inclined
disk with faint bar;
appears to contain several \HII\ regions scattered throughout the
disk.

{\it ESO 440-004:} nucleus?  bar traced by moderately
HSB region;
sprawling arms nearly perpendicular to the bar; possibly a merger
product.

{\it ESO 504-025:} possible faint nucleus; 2-step
disk; very diffuse outer
disk; appearance of small \HII\ regions scattered about disk; bulge?

{\it ESO 505-013:} =T1203-22; point-like nucleus;
well-defined,
faint spiral arms; possible weak bar.

{\it ESO 507-065:} very tiny, moderately HSB disk with
well-defined
apparent outer disk cutoff; entire disk outlined in ring of what appear
to be  \HII\ regions.

{\it ESO 443-079:} small, weak bar; very faint disk;
fairly symmetric disk but
with very asymmetric light distribution.

{\it ESO 443-083:} =T1310-22; dwarf companion present
in frame.

{\it ESO 444-002:} peculiar ``group''; Sm-like disk
surrounded by several
small, irregularly-shaped galaxy ``fragments''.

{\it ESO 444-033:} barred; very asymmetric light
distribution.
\end{flushleft}

\newpage

%figures 1a, b, and c
\figcaption{Azimuthally averaged V-band surface brightness profiles 
(corrected for inclination and Galactic extinction) versus radius, in 
arcseconds, for three of our extreme late-type spirals. These light 
profiles are based on full ellipse fits to the images. The solid lines 
are exponential fits to the inner portions of the disk.}

%figure 2
\figcaption{Our new B-band CCD magnitudes versus the photographic B magnitudes 
published for a number of our sample galaxies in the ESO Catalogue. The solid 
line is x=y.}

%figure 3a and b
\figcaption{V-band images of two of our extreme late-type spirals taken 
with the Curtis Schmidt telescope.
Panel a shows a 2500 second exposure of ESO~418-008 
($\sim1.4'\times1.4'$) with the outermost 
fitted ellipse from our fits to the 0.9m data overplotted. No excess 
light is visible outside of this ellipse in the Schmidt data. Panel~b 
shows ESO~305-009 ($\sim2.6'\times2.6'$) with the outermost fitted 
ellipse from our fits to the 1.5m data overplotted. A small amount of 
additional light is visible beyond this ellipse in the northeast 
and southwest corners of the image. This light totals less than $\sim$1\% 
of the total measured flux of the galaxy.}

%figure 4
\figcaption{\bmv\ color versus inclination for our extreme late-type 
spiral sample.}

%figure 5
\figcaption{\bmv\ color versus mean V surface brightness (corrected for 
Galactic extinction and inclination) for our 
extreme late-type spiral sample.}

%figure 6
\figcaption{\bmv\ color versus the logarithm of the ratio of the \HI\ 
mass to the V-band luminosity (corrected for Galactic extinction)
in solar units.}

%figure 7
\figcaption{Mean V surface brightness (corrected for Galactic 
extinction and inclination) versus the logarithm of the ratio of the 
\HI\ mass to the V luminosity (corrected for Galactic extinction) in 
solar units.}

%figure 8
\figcaption{Greyscale image of ESO 358-020, stretched to emphasize the 
abrupt change in surface brightness of the disk or surface brightness 
``step''.}

%figure 9
\figcaption{Major axis cut of ESO 358-020, plotted as pixel number 
versus the logarithm of the counts per pixel. A three pixel-wide strip 
along the galaxy major axis was sampled. The dotted line emphasizes 
the sharp change in slope in surface brightness (i.e. the surface 
brightness ``step'') which occurs in the light profile of this galaxy. 
The delineation of this feature becomes smeared out when the 
light profile is azimuthally averaged (see Figure 2b).}

%figure 10
\figcaption{Fundamental galaxy parameters (logarithm of the \HI\ mass; 
B - V; and absolute V magnitude) versus Hubble Type (as 
assigned from the present data set; see
Table~4) for our 
extreme late-type spiral sample.}

\figcaption{Plates: V-band images of our extreme late-type spiral 
galaxies. Plates 1--3 show our 0.9m data and Plates 4--7 are the 1.5m 
data. Each object frame was individually scaled to best emphasize the galaxy's 
morphology in the central regions.}

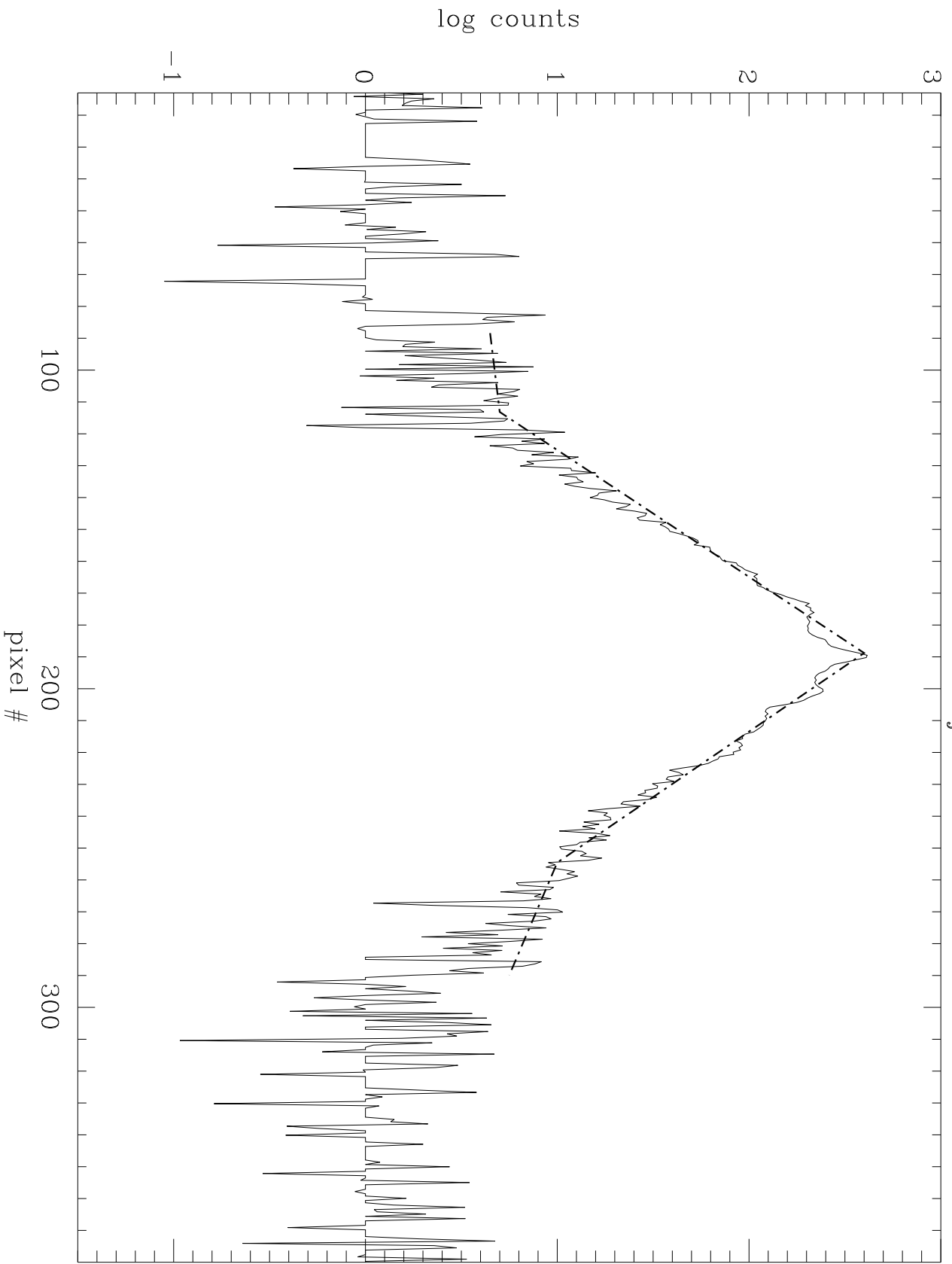
\end{document}

\begin{deluxetable}{lcccccccccccccc}
\tablewidth{60pc}
\newcommand{\p}{$\pm$}
\tablenum{3}
\pagestyle{empty}
\footnotesize
\tablecaption{Model-Free Photometric Parameters}

\tablehead{
\colhead{Galaxy Name}          & \colhead{m$_{B,25}$} &
\colhead{m$_{B}$}  &    
%\colhead{SB$_{B,out}$} &
\colhead{m$_{V,25}$} &  \colhead{m$_{V}$}      &
\colhead{Size ($'$)} &
\colhead{$\mu_{B,sky}$} & \colhead{$\mu_{V,sky}$} &
\colhead{$\bar\mu_{B}$} & \colhead{$\bar\mu_{V}$} &
\colhead{SB$_{V,outer}$}     & 
\colhead{a/b} & \colhead{PA} & \colhead{i} &
\colhead{Comments}  \\ [.2ex]
\colhead{(1)} & \colhead{(2)} & \colhead{(3)}& \colhead{(4)}&
\colhead{(5)}& \colhead{(6)}& \colhead{(7)}& \colhead{(8)}& 
\colhead{(9)}& 
\colhead{(10)}& \colhead{(11)}& \colhead{(12)}& 
\colhead{(13)}& \colhead{(14)}& \colhead{(15)}     }

\startdata
{\bf 0.9m Data:} & \phm{15.7} & \phm{15.42\p$^{.133}_{.151}$} &
\phm{14.9} & \phm{14.82\p$^{.136}_{.156}$} & \phm{1.68} & \phm{21.3} &
\phm{20.4} & \phm{24.93} & \phm{24.34} & \phm{25.4} & \phm{1.25} &
\phm{-40} & \phm{37.0} & \phm{A,e} \nl

ESO547-020 & 15.9 & 15.34\p.031 & 14.8 & 14.86\p.031 & 1.85 & 22.3 & 21.1 & 
24.99 & 24.52 & 26.0 & 1.33 & 45 & 41.7 & \nodata \nl

ESO418-008 &      13.9 & 13.92\p.015 & 13.5 & 13.51\p.010 & 
1.91 & 22.4 & 21.5 & 
23.36 & 22.96 & 26.6 & 1.72 & 140 & 55.0 & \nodata \nl

ESO418-009 &      14.3 & 14.18\p.025 & 13.7 & 13.67\p.013 & 1.68 & 22.4 & 21.6 & 
23.75 & 23.25 & 26.0 & 1.18 & 120 & 32.0 & \nodata \nl

ESO358-020 &      15.0 & 14.72\p.026 & 14.2 & 14.13\p.024 & 1.93 & 22.7 & 21.9 &
24.53 & 23.94 & 26.4 & 1.25 & 135 & 37.1 & \nodata \nl

ESO549-002 &      15.1 & 14.88\p$^{.190}_{.230}$ & 14.4 & 14.26\p.020 & 2.06 & 22.6 & 21.8 &
24.77 & 24.14 & 25.8 & 1.33 & 25 & 41.7 & a \nl

ESO359-016 &      15.2 & 15.18\p$^{.136}_{.156}$ & 14.7 & 14.64\p.016 & 2.02 & 22.7 & 21.9 &
23.91 & 23.37 & 27.3 & 3.70 & 50 & 75.4 & \nodata \nl

ESO359-031 &      14.9 & 14.78\p.032 & 14.3 & 14.22\p.013 & 1.37 & 22.0 & 21.0 &
23.98 & 23.42 & 25.6 & 1.11 & 75 & 26.0 & \nodata \nl

SGC0448-39 &      17.2 & 16.86\p$^{.240}_{.310}$ & 16.5 & 16.38\p.024 & 1.05 & 22.5 & 21.7 & 
24.78 & 24.30 & 26.6 & 2.13 & 82 & 63.4 & \nodata \nl

ESO422-005 &      15.1 & 14.98\p.170 & 14.8 & 14.59\p.019 & 1.70 & 22.6 & 21.8 & 
24.46 & 24.06 & 26.1 & 1.33 & 15 & 41.7 & \nodata \nl

ESO552-066 &      16.3 & 16.30\p.072 & 15.8 & 15.79\p.095 & 1.42 & 21.9 & 21.0 &
25.13 & 24.62 & 25.6 & 1.67 & 125 & 54.1 & e \nl

NGC2131 &      14.7 & 14.68\p.024 & 14.2 & 14.20\p.003 & 1.13 & 22.0 & 21.0 &
22.82 & 22.34 & 25.3 & 2.0 & 115 & 61.2 & \nodata \nl

ESO425-008 &      17.4 & 17.20\p.084 & 17.4 & 17.06\p.110 & 1.20 & 21.9 & 21.1 &
24.72 & 24.58 & 27.3 & 4.0 & 80 & 78.8 & \nodata \nl

AM0605-341 &      14.6 & 14.54\p.007 & 14.1 & 14.09\p.012 & 1.27 & 22.6 & 21.7 &
23.57 & 23.12 & 26.1 & 1.11 & 90 & 26.0 & e \nl

ESO497-007 &      16.8 & 16.38\p$^{.352}_{.531}$ & 16.2 & 15.90\p.034 & 1.48 & 22.4 & 21.6 &
25.62 & 25.14 & 26.0 & 1.25 & 135 & 37.4 & a \nl

ESO373-026 &   \nodata & \nodata & 13.1 & 13.15\p.021 & 2.22 & \nodata & 20.9 &
\nodata & 23.52 & 24.5 & 1.0 & 90 & 2.6 & a,b,c,h \nl

ESO500-032 &   \nodata & 13.83\p$^{.660}_{1.94}$ & \nodata & 
13.24\p$^{.653}_{1.85}$ & 1.76 & 21.4 & 20.9 & 23.45 & 22.86 & 23.3 & 
1.25 & 140 & 37.0 & d \nl

ESO377-017 &      17.3 & 17.0\p.056 & 16.4 & 16.33\p.040 & 1.56 & 22.2 & 21.3 &
25.20 & 24.54 & 27.1 & 3.57 & 30 & 76.5 & \nodata \nl

ESO503-022 &  14.9 & 14.90\p.021 & 14.3 & 14.31\p.004 & 1.53 & 21.0 & 20.5 & 
23.70 & 23.11 & 25.6 & 2.0 & 120 & 61.3 & \nodata \nl

ESO504-010 &      16.4 & 16.18\p.040 & 15.8 & 15.66\.03 & 1.44 & 22.4 & 21.6 & 
24.61 & 24.09 & 26.6 & 2.50 & 15 & 67.2 & \nodata \nl

ESO504-017 &      14.5 & 14.51\p.070 & 14.0 & 13.96\p.011 & 1.10 & 22.4 & 21.6 & 
23.18 & 22.63 & 25.9 & 1.16 & 103 & 30.8 & \nodata \nl

ESO440-039 &       15.6 & 15.40\p.041 & 15.0 & 14.80\p.010 & 1.70 & 22.3 & 21.6 &
24.75 & 24.15 & 26.5 & 1.49 & 113 & 48.8 & \nodata \nl

ESO440-049 & \nodata & \nodata & 13.9 & 13.89\p.034 & 2.22 & \nodata & 20.4 & 
\nodata & 24.08 & 25.8 & 1.18 & 60 & 32.0 & c,g \nl

ESO441-023 &       16.3 & 16.03\p.024 & 15.5 & 15.45\p.035 & 1.26 & 22.0 & 21.3 &
25.11 & 24.53 & 25.5 & 1.05 & 90 & 18.4 & e \nl

ESO380-025 &  \nodata & \nodata & 14.1 & 14.15\p.009 & 1.63 & \nodata & 20.3 & 
\nodata & 22.33 & 25.4 & 4.0 & 13 & 76.9 & c \nl

ESO443-080 &      14.1 & 14.18\p.033 & 13.8 & 13.78\p.005 & 1.89 & 21.2 & 20.7 & 
23.60 & 23.20 & 25.0 & 1.72 & 150 & 54.9 & \nodata \nl

ESO508-034 &      15.0 & 15.07\p.044 & 14.4 & 14.49\p.022 & 1.53 & 21.2 & 20.7 &
23.76 & 23.18 & 25.6 & 2.22 & 129 & 64.6 & \nodata \nl

ESO445-007 &      16.3 & 16.33\p.041 & 16.1 & 16.12\p.021 & 1.14 & 21.3 & 20.4 & 
24.26 & 24.05 & 26.1 & 2.50 & 0 & 68.2 & \nodata \nl

ESO510-026 &      15.7 & 15.42\p$^{.133}_{.151}$ & 14.9 & 
14.82\p$^{.136}_{.156}$ & 1.68 & 21.3 & 20.4 & 24.93 & 24.33 & 25.7 & 
1.25 & 130 & 37.0 & e \nl

ESO446-053 & \nodata & \nodata & 13.5 & 13.58\p.009 & 1.23 & \nodata & 19.1 & 
\nodata & 21.67 & 24.3 & 2.50 & 165 & 67.2 & c,e \nl

{\bf 1.5m Data:} & \phm{15.7} & \phm{15.42\p$^{.133}_{.151}$} & 
\phm{14.9} & \phm{14.82\p$^{.136}_{.156}$} & \phm{1.68} & \phm{21.3} & 
\phm{20.4} & \phm{24.93} & \phm{24.34} & \phm{25.4} & \phm{1.25} & 
\phm{-40} & \phm{37.0} & \phm{A,e} \nl

\tablebreak

ESO482-005 &      15.7 & 15.51\p.051* & 15.0 & 15.02\p.023* & 2.60 & 22.5 & 21.8 &
24.36 & 23.87 & 25.8 & 5.56 & 80 & 81.3 & \nodata \nl

ESO358-015 &      15.6 & 15.40\p.071* & 15.0 & 14.86\p.018* & 1.53 & 22.5 & 21.7 & 
24.54 & 24.00 & 26.0 & 1.45 & 180 & 46.7 & \nodata \nl

ESO548-050 &      15.0 & 14.96\p.016* & 14.5 & 14.42\p.008* & 2.92 & 22.4 & 21.6 &
24.28 & 23.73 & 25.4 & 4.54 & 179 & 78.4 & \nodata \nl

ESO358-060 &      15.8 & 15.76\p.023 & 15.8 & 15.58\p.034 & 2.24 & 22.2 & 21.5 &
24.28 & 24.10 & 25.6 & 5.56 & 102 & 85.2 & \nodata \nl

ESO359-029  &      14.9 & 14.91\p.035 & 14.3 & 14.29\p.024 & 1.92 & 22.3 & 21.5 &
24.33 & 23.71 & 25.0 & 1.78 & 25 & 56.4 & \nodata \nl

ESO305-009 &      14.5 & 13.73\p$^{.14}_{.17}$* & 13.2 & 13.11\p.053* & 4.63 & 22.6 & 21.8 &
25.15 & 24.53 & 25.4 & 1.64 & 50 & 52.9 & a \nl

ESO487-019 &      14.1 & 14.06\p.016* & 13.5 & 13.41\p.010* & 2.05 & 22.4 & 21.7 &
23.70 & 23.05 & 25.2 & 1.67 & 102 & 53.5 & \nodata \nl
%\tablebreak

ESO488-049 &   15.7 & 15.36\p.034* & 15.0 & 14.83\p.016* & 2.02 & 22.4 & 21.8 & 
24.68 & 24.15 & 25.5 & 2.17 & 132 & 63.1 & \nodata \nl

ESO431-015 & \nodata & \nodata & 16.4 & 16.38\p.051 & 1.40 & \nodata & 21.7 & 
\nodata & 24.0 & 24.6 & 5.0 & 128 & 83.0 & e,f,h \nl

ESO318-024 &      14.4 & 14.18\p.025* & 13.7 & 13.54\p.010* & 3.06 & 22.2 & 21.6 &
24.85 & 24.21 & 25.6 & 1.43 & 160 & 46.2 & a,f \nl

ESO502-016 &       14.8 & 14.70\p.032* & 14.2 & 14.12\p.018* & 2.46 & 22.1 & 21.3 &
24.29 & 23.71 & 26.1 & 2.50 & 79 & 67.1 & \nodata \nl

ESO438-005 &       15.0 & 14.96\p.029* & 14.6 & 14.52\p.048* & 3.40 & 22.1 & 21.4 &
24.50 & 24.05 & 25.3 & 5.0 & 60 & 80.0 & a \nl

ESO440-004 &       14.4 & 14.23\p$^{.235}_{.300}$ & 14.1 & 
13.90\p$^{.115}_{.129}$ & 3.73 & 22.1 & 21.3 & 24.78 & 24.45 & 25.6 & 
2.38 & 65 & 65.7 & a \nl

ESO504-025  &      15.0 & 14.95\p.056 & 14.6 & 14.34\p.021 & 2.42 & 22.2 & 21.4 &
25.22 & 24.61 & 26.0 & 1.30 & 20 & 39.9 & \nodata \nl

ESO505-013 &       13.2 & 13.30\p.082* & 12.8 & 12.82\p.075* & 2.96 & 22.0 & 21.3 & 
24.23 & 23.75 & 24.9 & 1.05 & 45 & 18.3 & a \nl

ESO507-065 &      15.9 & 15.94\p.073 & 15.5 & 15.51\p.029 & 1.26 & 22.1 & 21.3 &
24.01 & 23.58 & 24.2 & 2.63 & 12 & 69.6 & \nodata \nl

ESO443-079 &      15.7 & 15.46\p.019 & 15.1 & 15.01\p.033 & 2.18 & 22.1 & 21.4 &
24.51 & 24.06 & 26.0 & 3.23 & 180 & 72.8 & \nodata \nl

ESO443-083 & 13.2 & 13.31\p.006 & 12.8 & 12.76\p.003 & 3.70 & 21.9 & 21.1 
& 23.09 & 22.54 & 24.8& 4.76 & 152 & 79.0 & \nodata \nl

ESO444-002 &      15.4 & 15.16\p.012 & 14.8 & 14.75\p.012 & 1.45 & 22.2 & 21.4 &
24.41 & 24.01 & 25.5 & 1.18 & 83 & 32.2 & a,g \nl

ESO444-033 &      15.0 & 14.97\p.032 & 14.6 & 14.54\p.007 & 2.03 & 22.2 & 21.5 &
23.79 & 23.35 & 25.4 & 3.45 & 88 & 74.3 & \nodata \nl

\enddata

\tablecomments{Explanation of columns: (1) galaxy name; (2) estimated 
apparent B magnitude within 
the 25 mag arcsec$^{-2}$ isophote; (3) measured apparent blue 
aperture
magnitude; (4) estimated apparent V magnitude within the 25 
mag arcsec$^{-2}$ isophote; (5) measured apparent V aperture
magnitude; (6) diameter in arcminutes at the outermost measured 
isophote, measured from the V image; (7) \& (8) mean B and V sky 
surface brightnesses for image, in mag arcsec$^{-2}$; (9) \& (10) 
mean measured B and V surface brightnesses (no extinction or 
inclination corrections); (11) V surface brightness in magnitudes
arcsec$^{-2}$ 
of the last measured isophote; (12) axial ratio; (13) position angle, 
in degrees; (14) 
inclination, in degrees (see Text); (15) Comments. }

\tablenotetext{*} {asterisks denote cases where there may be a zero 
point uncertainty in both B and V of up to 0.24 magnitudes}
\tablenotetext{a} {many field stars present in galaxy aperture; 
photometric uncertainty likely to be larger than the formally quoted error}
\tablenotetext{b} {galaxy large compared with field of view; additional 
photometric uncertainty may result from poor sky background 
determination}
\tablenotetext{c} {no B data obtained}
\tablenotetext{d} {very bright star occults much of galaxy; large 
photometric uncertainty}
\tablenotetext{e} {possible stellar contamination in galaxy aperture}
\tablenotetext{f} {peculiarly shaped galaxy isophotes; inclination uncertain}
\tablenotetext{g} {photometry includes main galaxy only, not the 
possible 
companions (see Plates)}
\tablenotetext{h} {high Galactic extinction (see Table~1)}

\end{deluxetable}

\begin{deluxetable}{llccccccc}
\tablewidth{39pc}
\tablenum{4}
\tablecaption{Derived Optical Parameters}

\tablehead{
\colhead{Galaxy Name}          &
\colhead{Class}  &    
\colhead{M$_{B}$}      &
\colhead{M$_{V}$}     &    
\colhead{A$_{outer}$}  & \colhead{$\bar\alpha_{V}$}  &
\colhead{log$\frac{{\cal M}_{HI}}{L_{V}}$} & \colhead{$\bar\mu_{B,face-on}$} &
\colhead{$\bar\mu_{V,face-on}$}\\[.2ex]
\colhead{(1)} & \colhead{(2)} & \colhead{(3)} & \colhead{(4)} &
\colhead{(5)} & \colhead{(6)} & \colhead{(7)} & \colhead{(8)} &
\colhead{(9)}}

\startdata
{\bf 0.9m Data:} & \phm{SBdm} & \phm{-16.8} & \phm{-17.3} & \phm{14.5} 
& \phm{3.02} & \phm{0.24} & \phm{25.2} & \phm{24.8} \nl

ESO547-020 & SBdm & -16.8 & -17.3 & 14.5 & 3.02 & 0.24 & 25.2 & 24.8 \nl

ESO418-008 &      SBd & -17.0 & -17.4 & 8.33 & 0.87 & -0.21 & 24.0 & 
23.6 \nl

ESO418-009 &      SBdm & -16.2 & -16.7 & 5.75 & 0.69 & -0.50 & 23.9 & 
23.4 \nl

ESO358-020 &      SBdm & -17.1 & -17.7 & 12.8 & 1.42 & -1.08 & 24.8 & 
24.2  \nl

ESO549-002 &      SABdm & -16.1 & -16.6 & 8.8 & 2.39 & -0.66 & 25.0 & 
24.4 \nl

ESO359-016 &      SABd & -16.0 & -16.6 & 10.2 & 1.22 & 0.038 & 25.4 & 
24.9 \nl

ESO359-031 &      SBd & -16.6 & -17.1 & 7.42 & 1.06 & -0.47 & 24.1 & 
23.5 \nl

SGC0448-39 &      SABm & -15.4 & -15.8 & 8.55 & 1.45 & 0.67 & 25.6 & 
25.2 \nl

ESO422-005 &      Sdm & -16.1 & -16.5 & 8.38 & 1.20 & 0.40 & 24.8 & 
24.4 \nl

ESO552-066 &      Sm & -15.4 & -15.9 & 8.91 & 3.02 & 0.19 & 25.6 & 25.1 
\nl

NGC2131 &      BCD? & -16.8 & -17.3 & 6.56 & 0.87 & 0.17 & 23.6 & 23.1 \nl

ESO425-008 &      SBm: & -14.0 & -14.1 & 6.01 & 1.55 & 0.88 & 26.5 & 26.4 
\nl

AM0605-341 &      SBdm & -15.0 & -15.4 & 2.89 & 0.88 & -0.08 & 23.6 & 
23.2 \nl

ESO497-007 &      SABm & -16.8 & -16.9 & 15.0 & 4.66 & 0.18 & 25.2 & 
24.8 \nl

ESO373-026 &   SBc & \nodata & -19.9 & 21.1 & 3.92 & -0.38 & \nodata & 
23.0 \nl

ESO500-032 &      Scd: & -18.7 & -19.2 & 15.1 & 8.18 & -0.92 & 23.4 & 
22.9 \nl

ESO377-017 &      Sm & -15.6 & -16.1 & 13.3 & 4.59 & 0.94 & 26.5 & 25.9 
\nl

ESO503-022 &      SBm pec & -17.0 & -17.5 & 9.86 & 1.45 & -0.36 & 24.3 
& 23.7 \nl

ESO504-010 &      SBdm & -15.7 & -16.2 & 9.52 & 1.48 & 0.23 & 25.5 & 25.0 
\nl

ESO504-017 &      Scd/BCD & -17.5 & -18.0 & 7.25 & 0.61 & 0.095 & 22.9 
& 22.4 \nl

ESO440-039 &       SBm & -16.7 & -17.2 & 12.2 & 1.89 & -0.21 & 25.0 & 
24.4 \nl

ESO440-049 & Sc & \nodata & -18.5 & 18.1 & 2.34 & -0.39 & \nodata & 
24.0 \nl

ESO441-023 &       SABm & -16.5 & -17.0 & 10.9 & 2.82 & -0.023 & 24.8 & 
24.3 \nl

ESO380-025 &  SBdm & \nodata & -18.7 & 16.4 & 2.62 & -0.28 & \nodata & 
23.8 \nl

ESO443-080 &      Scd & -18.1 & -18.4 & 14.3 & 2.50 & 0.094 & 23.9 & 
23.6 \nl

ESO508-034 &      S pec & -17.0 & -17.5 & 10.4 & 1.49 & 0.30 & 24.4 & 
23.8 \nl

ESO445-007 &      SABm: & -15.1 & -15.3 & 6.38 & 1.65 & 0.60 & 25.2 & 
25.0 \nl

ESO510-026 &      SBcd & -17.2 & -17.7 & 14.7 & 2.49 & 0.31 & 24.9 & 
24.3 \nl

ESO446-053 & SABdm & \nodata & -17.6 & 5.92 & 0.97 & -0.41 & \nodata & 
22.4 \nl

{\bf 1.5m Data:} & \phm{SBdm} & \phm{-16.8} & \phm{-17.3} & \phm{14.5}
& \phm{3.02} & \phm{0.24} & \phm{22.4} & \phm{22.4} \nl

ESO482-005 &      SBd & -16.5 & -17.0 & 19.0 & 2.51 & 0.19 & 26.4 & 
25.9 \nl

ESO358-015 &      Sdm & -15.8 & -16.3 & 7.72 & 1.10 & -0.37 & 25.0 & 
24.4 \nl

ESO548-050 &      SABcd & -16.9 & -17.5 & 20.2 & 3.13 & 0.096 & 26.0 & 
25.4 \nl

ESO358-060 &      Sm & -14.1 & -14.3 & 6.13 & 1.19 & 0.68 & 27.0 &
26.8 \nl

ESO359-029  &      Sd & -15.1 & -15.8 & 5.71 & 1.0 & -0.38 & 25.0 & 
24.4 \nl

ESO305-009 &      SABc & -16.6 & -17.2 & 15.1 & 2.93 & 0.52 & 25.6 & 
25.0 \nl

ESO487-019 &      SBd & -17.8 & -18.4 & 13.7 & 1.74 & -0.47 & 24.2 & 
23.5 \nl

ESO488-049 &   SBcd & -16.4 & -16.9 & 12.8 & 1.78 & 0.52 & 25.5 & 24.9 \nl

ESO431-015 & Sm: & \nodata & -14.6 & 5.46 & 1.48 & 0.78 & \nodata & 
25.7 \nl

ESO318-024 &      S pec & -16.0 & -16.6 & 8.76 & 1.32 & 0.26 & 24.9 & 
24.4 \nl

ESO502-016 &       SBd & -16.6 & -17.1 & 12.2 & 1.38 & -0.44 & 25.1 & 
24.5 \nl

ESO438-005 &       SBd & -16.9 & -17.3 & 21.9 & 4.58 & 0.59 & 26.2 & 
25.8 \nl

ESO440-004 &       SBcd & -17.8 & -18.0 & 24.2 & 3.65 & 0.16 & 25.4 & 
25.1 \nl

ESO504-025  &      Sd & -16.7 & -17.2 & 13.6 & 2.05 & -0.063 & 25.2 & 
24.6 \nl

ESO505-013 &       SABc & -19.0 & -19.4 & 23.0 & 4.16 & 0.21 & 24.1 & 
23.6 \nl

ESO507-065 &      Sm pec & -16.9 & -17.2 & 12.0 & 3.27 & 0.29 & 24.8 & 
24.4 \nl

ESO443-079 &      SABd & -16.8 & -17.2 & 16.5 & 2.19 & 0.26 & 25.6 & 
25.2 \nl

ESO443-083 & Scd & -19.5 & -20.0 & 37.0 & 4.47 & -0.19 & 24.7 & 24.2 \nl

ESO444-002 &      Sm pec & -16.4 & -16.7 & 8.08 & 1.25 & 0.33 & 24.4 & 
24.0 \nl

ESO444-033 &      SBdm & -17.5 & -17.9 & 17.5 & 2.21 & -0.092 & 25.0 &
24.6 \nl

\enddata

\tablecomments{Explanation of columns: (1) galaxy name; (2) updated
Hubble classification based on new data; (3) absolute B magnitude, 
corrected for Galactic extinction using the extinctions and distances 
quoted in 
Table~1; (4) absolute V magnitude, corrected for Galactic extinction; 
(5) diameter, in kiloparsecs, at outermost measured isophote and corrected 
for Galactic extinction according to the RC3 (de Vaucouleurs 
{\it et al.} 1991); 
(6) mean 
disk scale in V-band in kiloparsecs (see Text); (7) logarithm 
of HI mass over V-band luminosity, in solar units; (8) Galactic 
extinction-corrected, mean B-band 
surface brightness in mags arcsec$^{-2}$
within the last measured galaxy isophote, corrected to 
a face-on value (see Text); 
(9) Galactic extinction-corrected, mean V-band surface brightness 
in mags arcsec$^{-2}$ within the last measured galaxy isophote, 
corrected to a face-on value.}

\end{deluxetable}